\documentclass[11pt,a4paper]{article} 
\pdfoutput=1
\usepackage{jcappub}
\usepackage{epsfig}
\usepackage{amssymb}
\usepackage{amsmath}
\usepackage{graphicx}
\usepackage{hyperref}

\def\rt{{\tilde r}}
\def\vu{{\bf u}}
\def\pmb#1{\setbox0=\hbox{#1}%
\kern-.025em\copy0\kern-\wd0
\kern.05em\copy0\kern-\wd0
\kern-.025em\raise.0433em\box0}

\def\vV{{\bf V}}

\def\kms{\, {\rm km }\, {\rm s}^{-1}}

\def\be{\begin{equation}}
\def\ee{\end{equation}}

\def\dd{{d}} 

\def\ka{\kappa_{_{\rm A}}}

\newcommand*\Laplace{\mathop{}\!\mathbin\bigtriangleup}

\begin{document}
\title{Testing Lorentz invariance of dark matter with satellite galaxies}

\author[a]{Dario Bettoni} \emailAdd{d.bettoni@thphys.uni-heidelberg.de}
\author[b]{Adi Nusser} \emailAdd{adi@physics.technion.ac.il}
\author[c]{Diego Blas} \emailAdd{diego.blas@cern.ch}
\author[c,d,e]{Sergey Sibiryakov} \emailAdd{sergey.sibiryakov@cern.ch}
\affiliation[a]{Institut f\"ur Theoretische Physik, Ruprecht-Karls-Universit\"at Heidelberg\\ Philosophenweg 16, 69120 Heidelberg, Germany}
\affiliation[b]{Physics Department and the Asher Space Science Institute - Technion, Haifa 32000, Israel}
\affiliation[c]{Theoretical Physics Department, CERN, CH-1211 Geneva 23, Switzerland}
\affiliation[d]{Institute of Physics, LPPC, Ecole Polytechnique
  F\'ed\'erale de Lausanne, CH-1015 Lausanne, Switzerland}
\affiliation[e]{Institute for Nuclear Research of the Russian Academy of Sciences,60th October Anniversary Prospect, 7a, 117312 Moscow, Russia}

\date{\today}
\abstract{
We develop the framework for testing Lorentz invariance  in the dark
matter sector using galactic dynamics. We consider a Lorentz violating
(LV) vector field acting on the dark matter component of a satellite
galaxy orbiting in a host halo. We introduce a numerical model for the
dynamics of satellites in a galactic halo and for a galaxy in a rich
cluster to explore observational consequences of such an LV field. The
orbital motion of a satellite excites a time dependent LV force which
greatly affects its internal dynamics. Our analysis points out key
observational signatures which serve as probes of LV
forces. These include modifications to the line of sight velocity
dispersion, mass profiles and shapes of satellites.  
With future data and a more detailed modeling these
signatures can be exploited to constrain a new region of the parameter
space describing  
the LV in the dark matter sector.  
}


\arxivnumber{1702.07726}
\maketitle

\section{Introduction}
\label{sec:int}
The evolution of the Universe and the observed cosmic structures is  well embraced  by the $\Lambda$CDM model \cite{Ade:2015xua}. The model is based on General Relativity (GR) with a cosmological constant term and on the matter content provided by Standard Model (SM)  plus a dark matter (DM) component. One of the  fundamental pillars of such construction is the assumption of Lorentz Invariance of the underlying theory. This requirement has far reaching consequences for both gravitational interaction and the construction of the SM.  Indeed, Lorentz symmetry is one of the most solid and tested symmetries, with 
extremely tight experimental  constraints  in both the gravitational
and particle physics sectors
\cite{Will:2014xja,Mattingly:2005re,Kostelecky:2008ts, 
Liberati:2012th,Liberati:2013xla,Berti:2015itd}. 

Certain theories  of quantum gravity involve some degree of Lorenz
violation (LV)
\cite{Kostelecky:1988zi,Mavromatos:2007xe,Horava:2009uw} which,
despite originating at high energy, may  have  significant
consequences on all scales \cite{Blas:2009yd}. The absence of Lorentz
Invariance is
also the basis for interesting proposals for dark energy
\cite{Blas:2011en,Audren:2013dwa} and inflation 
\cite{ArkaniHamed:2003uz,Ivanov:2014yla,Adshead:2016iix}. It is hence
worthwhile to explore the   observable signatures of LV on the
dynamics of the low energy sector of the theory.  
There is no unique recipe for   breaking Lorenz Invariance. We consider here  a specific, yet well-motivated \cite{Jacobson:2000xp,Horava:2009uw,Blas:2010hb}, choice where a vector degree of freedom takes a time-like non vanishing vacuum expectation value, effectively breaking the Lorentz group. 
The most general action for such an LV vector was derived in
\cite{Jacobson:2000xp} and its astrophysical implications have been
widely investigated \cite{Jacobson:2005bg,Shao:2013wga,Yagi:2013ava}
(for a recent review on the phenomenology of LV theories see
\cite{Blas:2014aca}). 

Once new degrees of freedom are present, it is natural to consider
their coupling to other particles, unless some mechanism prevents them
or make them negligible. The constraints on these interactions are
very strong for SM particles \cite{Kostelecky:2008ts} (see
\cite{Colladay:1998fq} for a general framework to parameterize
LV extension of SM).
However, there are currently few and less stringent constraints on couplings of LV fields to the DM sector. A first investigation in this direction was undertaken in \cite{Blas:2012vn,Audren:2014hza} where the theory is checked against  the large scale structures and cosmic Microwave Background radiation (CMB). The effect of LV  is twofold. First, it results in breaking of the Weak Equivalence Principle (WEP) 
by modifying  the inertial mass of DM particles without
an equal compensation in the gravitational mass. Second, it introduces a non-trivial velocity dependent interaction. 

 In this paper we explore consequences of LV  on smaller, non-linear,  scales.
As we shall see, small scales structures probe a completely different
range of the model parameters. Indeed, the effects introduced by the
LV vector are screened above a certain parameter-dependent scale
implying that there may be significant modifications on small scales
while the  evolution on large scales remains essentially intact
\cite{Audren:2014hza}.  
Hence, the investigation of non-linear structures opens up a window on
unexplored values of parameters.

Our aim is  to offer a broad assessment of the possible  effects of the coupling between DM and the LV vector on small scales. In particular, we will be interested in the consequences of such coupling on the dynamics of DM dominated satellite galaxies orbiting inside significantly more massive  host  halos.
 Using this type of systems, we will define characteristic observable
 features that can  be used to constrain the LV models.  Satellites and galaxies
 have been exploited in the literature for testing breaking of the WEP
 in the dark sector
 \cite{Frieman:1993fv,Kesden:2006zb,Keselman:2009nx,Mohapi:2015gua}. Although the LV
 theory we consider here also generates WEP violation, the main focus
 is on specific aspects of LV. 

A full N-body simulation  would be  required in order to assess the exact level of LV effects in realistic scenarios. This is a formidable numerical effort which is beyond the scope of the current paper.
Instead we resort to a  semi-analytic approach which describes idealized situations.
The approach helps preparing  the ground for a more complex analysis, by pointing 
to promising paths for further explorations.

The paper is organized as follows. In section \ref{sec:LVDM} we introduce the model and we recall some previous results while in section \ref{sec:LVDMSSR} we discuss the weak field regime  which serves as a basis for the subsequent discussion. In section \ref{sec:LV_Hal} we focus in the implications
of LV for DM halos. Section \ref{sec:EOM_sat} describes the relevant equations for DM particles, while Section \ref{sec:LVDM_AR} presents the solutions of the LV field in the halos and useful insights into the physics of Lorentz-violating dark matter (LVDM). In section \ref{sec:numerical_results} we present the results of the numerical integration of the dynamics and, finally, in section \ref{sec:conclusions} we summarize the observational tests for the LVDM model and draw our conclusions.
Appendix \ref{app:analytic_LV} contains some analytic solutions.

\section{Lorentz violating dark matter}
\label{sec:LVDM}

We will assume that the Universe is endowed with a preferred time
direction or foliation of space-time that breaks Lorentz
invariance. This can be described by introducing the `Aether' vector $U^\mu$
 and forcing it to have a time-like unit norm that selects the
 preferred time direction. The most general low-energy action that
 describes the dynamics of this theory is the so called
 Einstein--Aether ($\AE$) action
 \cite{Jacobson:2000xp,Jacobson:2008aj} and is given by 
\begin{equation}\label{eq:AE_action}
S_{\textrm{IR}}=-\frac{M_0^2}{2}\int d^4x \sqrt{-g}\big(R+K^{\alpha\beta}{}_{\mu\nu}\nabla_\alpha U^\mu\nabla_\beta U^\nu +\lambda(U^\mu U_\mu-1)\big)\,,
\end{equation}
where $\lambda$ is a Lagrangian multiplier enforcing the unit norm of
the vector $U^\mu$, $M_0$ is a scale related to the Planck mass
\cite{Blas:2014aca} and 
\begin{equation}
K^{\alpha\beta}{}_{\mu\nu}= c_1 g^{\alpha\beta} g_{\mu\nu} +c_2 \delta^\alpha_\mu\delta^\beta_\nu +c_3\delta^\alpha_\nu\delta^\beta_\mu +c_4 U^\alpha U^\beta g_{\mu\nu}\,.
\end{equation}
The dimensionless parameters $c_i$ characterize the interaction between the LV vector and gravity. The strongest constraints on  these couplings come from Solar system physics through the post-Newtonian (PPN) parameters \cite{Will:2014xja}
$\alpha_1^\text{PPN}\le10^{-4}$ and $  \alpha_2^\text{PPN}\le 4\cdot 10^{-7}$,
which generically  imply \cite{Foster:2005dk}
$
|c_i| < 10^{-7}.
$
However, the Solar System bounds can be satisfied by specific
combinations of the couplings, which still yield part of the parameter
space totally unconstrained. The latter 
can be explored through other observations, as Big Bang
Nucleosynthesis (BBN) \cite{Carroll:2004ai}, dynamics of  binary
systems \cite{Shao:2013wga,Yagi:2013ava}, or linear cosmology
\cite{Audren:2014hza}. 

The bounds on possible direct couplings of SM to the  $\AE$-field are
very tight \cite{Kostelecky:2008ts,Liberati:2013xla} and can be safely
assumed to be zero for astrophysical and cosmological
implications. Hence, broken Lorentz symmetry will affect ordinary
matter only via gravitational interaction.  
This assumption can be relaxed in the case of DM. Since the relation
between DM and SM particles remains uncertain, there is no solid
reason to expect that their couplings to the LV sector may be of the
same order. From this point of view, the study of LV in the DM sector
offers a new handle in testing one of the most fundamental paradigms
of modern physics, with important consequences for quantum gravity.  

We thus equip the $\AE$ theory with an explicit coupling between the LV vector and DM which will break the Lorentz invariance of this component. This is achieved by modifying the action for a DM particle as follows \cite{Blas:2012vn}
\begin{equation}\label{eq:LVDM_Action}
S_{\textrm{dm}}=-m\int ds F(U^\mu {\cal V}_\mu/c)\,,
\end{equation}
where
$
ds\equiv \sqrt{g_{\mu\nu} dx^\mu dx^\nu}\,,
$
$
{\cal V}^\mu \equiv {dx^\mu}/{ds}
$
is the particle's four-velocity and  $F$ is an arbitrary positive
function normalized in such a way that $F(1)=1$. 
  
The sum of the two actions \eqref{eq:AE_action} and
\eqref{eq:LVDM_Action}, plus any other minimally coupled matter
species, defines the LVDM model whose small scales dynamics will be
now investigated.

\section{The weak-field and non-relativistic limit}\label{sec:LVDMSSR}

The dynamics of DM halos can be accurately described by the weak-field and non-relativistic limit of the action presented in the previous section. 
This is achieved by first expanding the action \eqref{eq:AE_action} together with \eqref{eq:LVDM_Action} around flat space to second order in the fields. We  decompose the metric as 
$g_{00}=1+2 \phi$, $ g_{i0}=0$ and $g_{ij}=-\delta_{ij}(1-2\psi)$ and the vector field as $U^\mu=\delta_{\mu 0}+u^\mu$.  Assuming   non-relativistic  velocities for the particles and that the time derivatives of the potentials are sub-leading, the weak field action reads
\begin{equation} \label{eq:PN_Action}
S=\frac{M_{\textrm{0}}^2}{2}\int d^4x\left(4\phi\nabla^2\psi-2\psi\nabla^2\psi+\kappa_A {\bf u}\cdot\nabla^2 {\bf u}+\kappa_B(\partial_i u^i)^2\right)
+\int d^4x\, \rho\left[\frac{\vV^2}{2 c^2}-\phi-\frac{Y}{2}\left({\bf u}-\frac{\vV}{c}\right)^2\right]\,,
\end{equation}
where we defined $Y=F'(1)$,  $\kappa_A\equiv c_1$ and $\kappa_B\equiv c_2+c_3$; and introduced $V^i=dx^i/dt$ and the DM density $\rho = m \sum_A \delta^{(3)}(x-x_A)$.
Notice that this action is invariant under Galilean
transformations that act on the velocity and Aether by a constant
shift 
\[
{\bf V}\mapsto {\bf V}+{\bf v}_0~,~~~~
{\bf u}\mapsto {\bf u}+\frac{{\bf v}_0}{c}\;.
\]
In what follows we mostly focus on the choice of
parameters with $\kappa_B=0$. This is a well defined effective theory
which simplifies our analysis and still yields an interesting
phenomenology. It is also favored by some theoretical arguments
\cite{Marakulin:2016net}. The most general case will be considered in
future work.  

In the special case of negligible Aether fluctuations  ($u^i=0$) the DM action is
\begin{equation}
\label{eq:pp_screened}
S_{pp}=\int d^4x\,\rho \left[\left(1-Y\right)\frac{\vV^2}{2}-\phi \right]\,,
\end{equation}
which shows that the inertial and gravitational masses are not
equivalent anymore and the Equivalence Principle is not
satisfied. This introduces a segregation between DM and baryons which in our
scenario  are not affected by LV\footnote{In fact it is enough that they have
different $Y$ parameter; we assume $Y_b=0$.}. 
The last term in the action \eqref{eq:PN_Action} gives rise to a
quadratic potential for the Aether inside regions with non-zero DM
density. 
The stability of the
Minkowski background requires this potential and the effective
inertial mass of DM particles 
to be positive, which implies
$0<Y<1$ \citep{Blas:2012vn}. 
The potential tends to align 
the Aether vector ${\textbf u}$ 
with the DM velocity leading to a suppression of LV effects 
on the dynamics (screening mechanism).  

Variations of \eqref{eq:PN_Action} with respect to the matter velocity
$\textbf{V}$, the LV vector $\textbf{u}$ 
and the gravitational potentials $\phi$ and $\psi$ yield the equations
of motion, 
\begin{equation}
\label{eq:v_gen}
\frac{1-Y}{c^2}\frac{dV^i}{ dt}+\partial_i\phi +\frac{Y}{c}\left(\frac{du^i}{dt}+(cu^j-V^j)\partial_i u^j\right)=0\,,
\end{equation}
\begin{equation}
\label{eq:u_gen}
\Laplace u^i=  \frac{8\pi GY\rho}{c^2 \ka}\left(u^i-\frac{\overline  V^i}{c}\right)\,,
\end{equation}
\begin{equation}\label{eq:Poisson_PN}
\Laplace\phi=\frac{4\pi G}{c^2}\rho\,,  \quad \quad \psi=\phi\,.
\end{equation}
Since  DM is collisionless, in the derivation 
of \eqref{eq:u_gen} we allow for  multiple streams 
each with its own  velocity  existing at a certain point. Therefore, the equation involves the  (mass weighted) mean particle velocity $\overline V^i$. 
Equation \eqref{eq:Poisson_PN} is the standard Poisson equation,
meaning that  
 the LV vector $\textbf{u}$ does not modify the relation 
between the gravitational potential and the mass density.
In addition, DM obeys mass conservation as described by the usual
continuity  equation.

\section{Lorentz violation in dark matter halos}\label{sec:LV_Hal}

Internal dynamics of DM halos 
are a  plausible testbed  of the direct coupling of the  LV field
to DM. Inside the virial radius of 
a halo, matter   is in 
near dynamical equilibrium and overlapping streams of particles moves
in random directions.   Thus, in the reference frame moving with the  
bulk motion of the halo,  the  mean particle velocity, $\overline
{\textbf{V}}$, is very close to zero.  
Since $\textbf{u}$ is sourced  by $\overline{\textbf{V}}$  through
\eqref{eq:u_gen}, an individual particle moving with a random velocity
$\textbf{V}\ne \overline {\textbf{V}}$ will always experience an LV
force through the term $c\textbf{u}-\textbf{V}$ in \eqref{eq:v_gen}. 

Still, the internal dynamics of an isolated halo moving with a
\textit{constant} bulk velocity is not significantly affected by LV.  
To see this, let us perform a Galilean boost to the frame moving with
the halo. In this frame the LV vector is time-independent.
Hence, the action for the DM particles
does not involve any
terms which depend explicitly on time, resulting in an ``energy-like''
conserved quantity.  
This can be demonstrated by multiplying the velocity equation
\eqref{eq:v_gen} by $\vV$ to obtain  
\begin{equation}
\frac{1-Y}{2c^2}\frac{\dd \vV^2}{\dd t}=-\vV \cdot {\bf \nabla} \left(\phi+\frac{Y}{2}{\bf u}^2\right)\, .
\end{equation}
This equation implies that the quantity 
\begin{equation}\label{eq:gen_en_cons}
{\cal E}=\frac{1-Y}{2c^2}\vV^2+\phi+\frac{Y}{2} {\bf u}^2
\end{equation}
is an integral of motion. 
Although the LV force still affects the halo dynamics, this
conservation law prevents any significant departures from the standard
evolution governed  
by gravity alone. 

\begin{figure}[!ht]
\centering
\includegraphics[scale=.25]{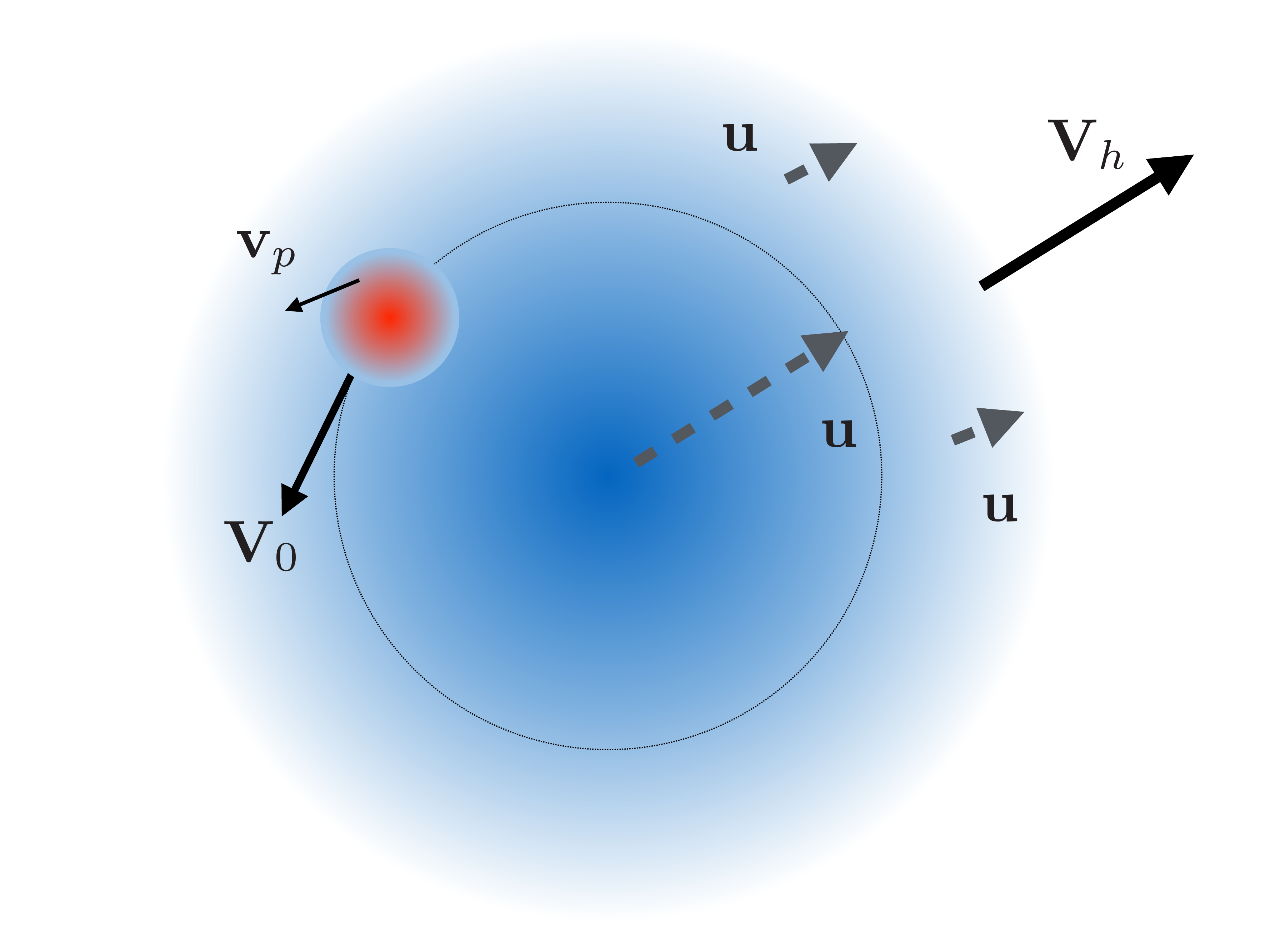}
\caption{Schematic representation of the satellite and the host halo
  system. 
A spherical host halo is moving with constant velocity ${\bf V}_h$
relative to the CMB frame.  ${\bf V}_0={\bf V}_s-{\bf V}_h$ 
is the bulk velocity of the satellite (subhalo) relative to the host
halo. ${\bf v}_p={\bf V}-{\bf V}_s$ is the relative velocity of a
particle  with respect to the subhalo velocity. 
The vector ${\bf u}$  vanishes outside of the host halo (it vanishes
in  the CMB frame) and is aligned with ${\bf V}_h$ at small scales due
to a screening mechanism  (see \eqref{eq:u_gen} and
\eqref{eq:sol_p}).}
\label{fig:halos} 
\end{figure}

The time independence is broken for satellite halos orbiting inside a more massive 
host halo.  
The orbital motion of these satellites 
introduces a time dependent LV vector  which 
breaks the conditions for the conserved quantity above. 
We will consider two types of systems. The first one is intended to mimic 
a dwarf satellite galaxy  in the halo of  the  Milky Way (MW). 
Most of these satellites are DM dominated, with baryons having only a
modest effect on their overall dynamical evolution.
The second system  is a galaxy in a massive  cluster. This system is
three orders of magnitude more massive than the first galactic system and
probes a different velocity regime of the LV force (see
below). We work with an  idealized configuration as summarized
schematically in  Fig.~\ref{fig:halos}.

\subsection{Equations of motion in the satellite frame of reference }
\label{sec:EOM_sat}

We write the equations of motion of particles belonging to a satellite
bound to a host halo as shown in Fig.~\ref{fig:halos}.  
The distance of a particle from the center of the satellite is denoted by $r$.
Furthermore, let us denote by $M_v$  and $r_v$ the virial mass and virial
radius of the satellite (the latter is defined as the radius inside
which the average density is $200$ times the background value). 
The actual mass and radius of the satellite will be denoted by $M_s$ and $r_s$,
respectively. Due to tidal disruption by the gravitational field of
the parent halo, $r_s$ is smaller than $r_v$.  
We also define $V_c=(G M_s/r_s)^{1/2}$ as  the circular velocity of a
baryonic tracer  moving  
in a circular  orbit at $r=r_s$ under the action of gravity alone.

Substituting in  \eqref{eq:v_gen}  ${\mathbf
  v}_p=\textbf{V}-\textbf{V}_s$ for the velocity of a particle in a
frame of reference moving with velocity $\textbf{V}_s$ with respect to
the CMB frame yields the modified Euler equation, 
\begin{equation}\label{eq:v_p}
(1-Y)\frac{d\tilde v^i_p}{d\tau}+\frac{c^2}{ V_c^2}\tilde\partial_i\phi_{sat} + \left(\frac{d\tilde V_s^i}{d\tau}+\frac{c^2}{V_c^2}\tilde\partial_i\phi_{h}\right)+Y\left[\frac{\dd(\tilde u^i-\tilde V^i_s)}{\dd\tau}+\left(\tilde u^j-\tilde V_s^j-\tilde v_p^j\right)\tilde \partial_i\tilde u^j\right]=0\,,
\end{equation}
where $\phi_h$ and $\phi_{sat}$ are, respectively,  the gravitational
potentials generated by the host and  satellite halos. We  have 
introduced the rescaled variables
\begin{equation}
\label{eq:rescaled}
\tilde x^i=x^i/r_{sc}
\,, \qquad \tau=t V_c/r_{sc}\,,\qquad \tilde u^i= c u^i/V_c\,,\qquad \tilde V^i=V^i/V_c\,,\qquad\tilde\partial_i=\partial/\partial \tilde x^i=r_{sc}\partial/\partial x^i\, .
\end{equation}
Here
\begin{equation}
r_{sc}=\sqrt{\frac{c^2\ka}{8 \pi G(1+\delta_v)Y\bar \rho}}=1.43\,{\rm Gpc}\,\sqrt{\left(\frac{10}{1+\delta_v}\right)\left(\frac{0.3}{\Omega_0}\right)\left(\frac{\ka}{Y}\right)}\frac{70\, (\rm{km}/\rm{s})/\rm{Mpc}}{H_0}\, ,
\label{eq:def_rsc}
\end{equation}
where\footnote{The precise value of $1+\delta_v$ depends on the halo
  profile. It is equal to $200/3$ for the halo density $\rho\propto
  r^{-2}$.} 
$1+\delta_v\approx 200/3$ is the DM overdensity at the virial
radius, 
$\bar \rho=3H_0^2\Omega_0/(8\pi G)$ is the average DM density in
the Universe, $H_0$ is the Hubble constant and  $\Omega_0$ is the
present day DM background density parameter. 
As we will see shortly,  $r_{sc}$ marks the screening length, above
which the LV force is suppressed. 

$\textbf{V}_s$ is identified as the velocity of the
satellite's central particle ($r=0$) and we consider parameters for
which ${\bf u}$ is aligned with ${\bf V_s}$ 
at the core of the satellite so that the LV force on the central
particle 
vanishes (see discussion in appendix
\ref{app:analytic_LV}). 
Then the central particle follows the same trajectory as it would have
under the influence of gravity alone. Moreover, we will neglect
the tidal gravitational forces exerted on
the satellite by the host halo, as we want to isolate the effect of
the LV interactions. In this approximation
the combination in the round brackets in
\eqref{eq:v_p} is zero and we obtain, 
\begin{equation}\label{eq:sat_p}
(1-Y)\frac{\dd \tilde v^i_p}{\dd \tau}=-\frac{c^2}{V_c^2}\tilde \partial^i\phi_{sat}-Y \left[ \frac{\dd  (\tilde u^i-\tilde V^i_s)}{\dd \tau} +( \tilde u^j-\tilde V^j_s-\tilde v_p^j)\tilde\partial_i  \tilde u^j\right]\,.
\end{equation}
This form of the equation will be used in numerical simulations.

\subsection{Solution to the Aether equation}
\label{sec:LVDM_AR}

We turn now to equation \eqref{eq:u_gen} which determines the LV vector $\textbf{u}$ from 
the density and the velocity fields. We consider cases where the radius of the satellite and the screening length are much smaller than the radius of the host halo.  
Therefore, as far as the internal dynamics of the satellite is concerned, we 
can treat the host halo as an infinite medium with constant density. 
We also assume that the satellite is spherical. 
Equation \eqref{eq:u_gen} can be rewritten in a dimensionless form,
\begin{equation}\label{eq:ueq_PN}
\tilde{\Laplace} \tilde u^i=  \frac{\rho(r)}{\rho(r_v)}(\tilde u^i-
{\tilde {\overline V}}^i)\,,
\end{equation}
where  $\rho$ is 
the mass density profile and $\rho(r_v)$ is the mass density of the satellite at the virial radius, cf.~\eqref{eq:def_rsc}.
Given that the Poisson equation \eqref{eq:Poisson_PN} is linearly
sourced only by the matter density, it can be easily integrated.
Note that if the typical variation of the density occurs on scales
larger than $r_{sc}$, equation \eqref{eq:ueq_PN} implies
$\tilde{\mathbf u}= \tilde{\bar{\mathbf V}}$. Then, in the case of
a single stream, i.e. ${\bf v}_p=0$, equation \eqref{eq:v_p} would
reduce to the standard Euler equation with the LV force 
screened away \cite{Blas:2012vn}. In the general case of multiple
streams, particles that do not move with the average velocity are
affected by the LV force. Still, the Euler equation is recovered upon
averaging over the whole set of particles.
 
We can now solve equation \eqref{eq:ueq_PN} in the two cases that are
relevant for our investigation. First, we will solve it for a single
halo with a time independent  velocity. The aim in this case is to
find the conditions on the parameters $r_{sc}$ for which the inner
part of such halo is unaffected by the LV force, i.e. the situation
where screening is efficient. Secondly, we will study what happens in
the case of a satellite orbiting around its host halo in the case of
screening happening in the latter. A detailed derivation of the
solutions is provided in appendix \ref{app:analytic_LV}.

\subsubsection{Solution for a single halo}
\label{sec:first}

Consider the idealized situation of a DM halo, described as a sphere
of constant density,  moving with a constant bulk velocity $\vV_h$ with respect to the CMB frame. Recall that for any
virialized object the average velocities of the particles inside it is
zero. In this case the equation \eqref{eq:ueq_PN} reads,
\begin{eqnarray}\label{eq:eq_u_parent}
\tilde{\Delta}\tilde u^i  = (\tilde u^i -\tilde V_h^i ) & \qquad & \text{for $r\le r_{h}$}\,,\\
\tilde{\Delta}\tilde u^i  =  \alpha^2 \tilde u^i& \qquad &  \text{for $r\ge r_{h}$}\,,
\end{eqnarray}
where $r_h$ is the physical radius of the halo and $\alpha^2 =
\rho_\text{out}/\rho_\text{in}$ is the ratio between the density
outside and inside the halo.
Note that in this case $\rho_\text{in}$ coincides with the virial
density of the halo. 
We take the solution to be continuous at $r=r_h$
together with its first derivatives.
We obtain that Aether vector is directed
along ${\vV}_h$ and has the following radial dependence
\be
\label{eq:sol_p}
\begin{split}
{\bf u}_\text{in}(r)&= \frac{{\vV}_h}{c}\left[1-\frac{(r_h+r_0) \sinh\left(r/r_{sc}\right)}{\sinh\left(r_h/r_{sc}\right)r}\right]\,,\\
{\bf u}_\text{out}(r)&= -\frac{{\vV}_h}{c} \frac{r_0}{r}e^{-\alpha(r-r_h)/r_{sc}}\,,
\end{split}
\ee
where $r_0$ is a constant fixed by matching the solution at $r=r_h$
and we have restored physical units. 

The solution \eqref{eq:sol_p} has three regimes, depending on the
ratio $r_h/r_{sc}$:  
\begin{figure}[!ht]
\centering
\includegraphics[scale=0.3]{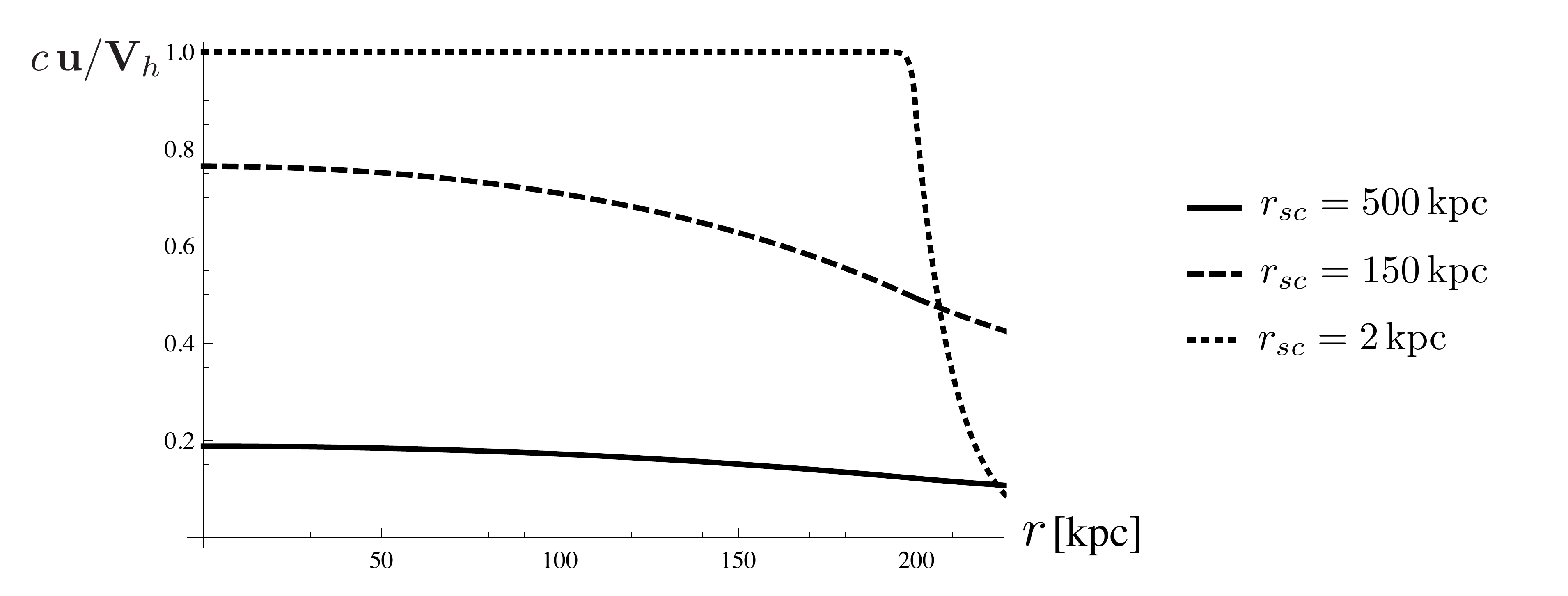}
\caption{Solution of the inner region for various values of $r_{sc}$
  of $2$ kpc, $150$ kpc and $500$ Mpc. The radius of the halo is taken
  to be $r_h=200$ kpc, corresponding to the virial radius of the Milky
  Way while $\alpha=0.1$.}
\label{fig:external_vector_solution}
\end{figure}
If $r_h/r_{sc} \gg 1$,  the variation of $u$ occurs abruptly close to
the edge of the halo (very efficient alignment). In  this case,
only a thin shell experiences the LV force.  
Throughout the interior of the halo ($r<r_h$), the effects of LV are almost
vanishing because $\textbf{u}\rightarrow \textbf{V}_h/c=const$.   
In the case $r_h/r_{sc} \ll 1$ the variation of the LV vector ${\textbf
  u}$ over the size of the halo is very small and the force given
by the last term in equation \eqref{eq:v_gen} is negligible. Thus, the
only consequence of LV in this case is the difference between the inertial and
gravitational masses. 
Finally, the velocity-dependent LV 
force is relevant in the cases where $r_{sc}\sim r_{h}$,
which means that particles inside the halo will experience a force
which depends on their position. 
Figure \ref{fig:external_vector_solution} shows  solutions for the three cases.

For our investigation of  the dynamics of the subhalo we will focus on
the case $r_{sc} \ll r_h$, for which  the LV effects of the host halo  
are screened and can be ignored. In
Fig.~\ref{fig:external_vector_solution} 
we see that  $r_{sc} \le 0.01\, r_h$ is sufficiently small to
guarantee a negligible LV force inside the halo.  
 For an MW-like halo, this  corresponds to  $r_{sc} \sim 1 - 10$ kpc
 and the LV effects of the external halo can be ignored  as long as
 the subhalo is 
 not  close to the edge. For example, the Draco dwarf is at a distance
 of $r_\text{Draco} \sim 80$ kpc, whereas the virial radius of the MW
 is $r_\text{MW} \sim 200$ kpc \cite{Dehnen:2006cm}. We find that
 at Draco's distance, even a value as high as  $r_{sc}\sim 50$ kpc
 suppresses LV effects of the parent halo.

\subsubsection{A satellite in a halo}
\label{sec:solving_u_satellite}

We now turn  to the dynamics of a DM satellite orbiting in its host halo. The 
host is assumed to have a constant velocity $\textbf{V}_h$ with
respect to the CMB and the satellite to move with velocity
$\textbf{V}_0=\textbf{V}_s-\textbf{V}_h$ with respect to the host halo
(see Fig.~\ref{fig:halos}). 
We work in a frame of reference whose origin,  $r=0$,  coincides with
the center of the satellite. 
The vector $\vu$ is the solution to the general equation
\eqref{eq:ueq_PN} with the following source,  
\begin{eqnarray}
\nonumber \vV( r)&=&\vV_s \qquad \text{for}\quad  r\le  r_s\,,\\
\vV( r)&=&\vV_h \qquad \text{for} \quad r> r_s\,,
\end{eqnarray}
where $r_s$ is the radius of the satellite. We assume that 
the satellite is spherically symmetric with $\rho\propto r^{-2}$ (see
appendix \ref{app:analytic_LV} for the solution in the case of a 
  more general density profile containing a core).  
In terms of the scaled variables \eqref{eq:rescaled}
the Aether equation reads,
 \begin{eqnarray}
  \label{eq:in}
\tilde{\Laplace} \tilde u^i&=& \left(\frac{\rt_v}{\rt}\right)^2(\tilde u^i -\tilde V^i_s) \qquad \text{for} \quad\rt\le \rt_s\,,\\
 \label{eq:out}
\tilde{\Laplace} \tilde u^i&=&\alpha_h^2(\tilde u^i-\tilde V^i_h)\qquad \text{for} \quad \rt> \rt_s\, , 
\end{eqnarray} 
where $\alpha_h^2= \rho_h/\rho_s(r_v)$ is the ratio between the
halo density and the virial density of the satellite. Notice that we
assume a  host halo of constant density.
The solution to the previous system of equations is
\begin{align}
\label{eq:genv}
\tilde\vu(\rt)-\tilde\vV_s&=
\frac{\tilde\vV_h-\tilde\vV_s}{1+\frac{n}{1+\alpha_h\rt_s}}
\left(\frac{\rt}{\rt_s}\right)^n \equiv -\tilde\vV_0 w_{in}(\tilde
r) \qquad
\text{for}\quad\tilde r\leq \tilde r_s\, ,\\
\label{eq:genoutv}
\tilde\vu(\rt)-\tilde\vV_s&=(\tilde\vV_h-\tilde\vV_s)\bigg[1-
\frac{n}{n+1+\alpha_h\rt_s}\,\frac{\rt_s}{\rt}\,
{\rm e}^{-\alpha_h(\rt-\rt_s)}\bigg]\equiv -\vV_0 w_{out}(\tilde r)
\qquad
\text{for}\quad\tilde r> \tilde r_s\, .
\end{align}
with 
\be
n=\frac{1}{2}\left(-1+\sqrt{1+4\tilde r_v^2}\right)\,. \label{eq:n}
\ee
This solution will be used as an input in the equation of motion \eqref{eq:sat_p}.

\subsubsection{Comparison with gravitational forces}

To obtain an order of magnitude estimate of the strength of the LV
effects, we  compare the time dependent part of the LV force with the standard
gravitational interactions, namely the gravitational force generated by the
satellite halo itself and the tidal effects produced by the
gravitational field of the host halo. From the equation
\eqref{eq:sat_p} and using the results of the previous section it can
be shown that the absolute value of the time dependent force produced by the
LV coupling on a particle belonging to a satellite on circular orbit
is
\begin{equation}
F_{\dot u}\simeq Yw( r) \frac{V_0^2}{R_0}\,,
\end{equation}
where $V_0$ is the orbital velocity of the satellite in the host halo,
$R_0$ is the distance from the center of the host and the function
$w(r)\leq 1$ has been defined in \eqref{eq:genv}, \eqref{eq:genoutv} and is shown for some cases in appendix \ref{app:analytic_LV}. 
The gravitational force generated by the satellite on a particle  is given by
\begin{equation}
F_{sat}=\frac{GM(r)}{r^2}=V^2_c \mathcal{M}\frac{ r_s}{ r^2}\,,
\end{equation}
where $V_c^2=GM_s/r_s$ and $\mathcal{M}=M(r)/M_s$. Hence,
\begin{equation}
\frac{F_{\dot u}}{F_{sat}}= Y\, w(r) \left(\frac{V_0}{V_c}\right)^2\frac{M_s}{M(r)}\frac{r^2}{r_s R_0}\,.
\end{equation}
Let us take  typical values  $V_0\sim 200$ $\kms$ and $V_c\sim 20$
$\kms$,  $r_s\sim 2$ kpc and  $R_0=100$ kpc. With these numbers the
ratio between the two forces at $r=r_s$ is  
\begin{equation}
\frac{F_{\dot u}}{F_{sat}}\sim 2\, Y\,w(r_s)\,.
\end{equation}
The ranges of the LV parameters studied in this
paper are listed in Table~\ref{table:param}. We see that for these ranges of parameters 
there will be an extra force acting on the particles in the 
satellite 
comparable to the gravitational force. 
The effect is maximal at the edge of the satellite halo given that
$|w(r)|\leq 1$ is a monotonically growing function, as can be see from
figures \ref{fig:cored_vs_rho2} and 
\ref{fig:cored_vs_rho2_screen} in the Appendix. 

Let us also consider an order of magnitude estimate of the
gravitational tidal force of the host halo.  
The force exerted by the host galaxy onto a particle that lies at the
boundary of the satellite's halo (on the side facing the galactic
center) is given by 
\begin{equation}
F_{tidal}=\frac{GM_h}{(R_0-r_s)^2}-\frac{GM_h}{R_0^2}\sim V^2_0\frac{r_s}{R_0^2},
\end{equation}
where we have used the fact that $V_0=\sqrt{GM_h/R_0}$.
Hence, the ratio between the two forces is
\begin{equation}
\frac{F_{\dot u}}{F_{tidal}} = Yw( r_s)\frac{R_0}{r_s}
\,.
\end{equation}
If we insert characteristic values in the previous equation we find
that at the edge of the satellite the force due to the LV vector can
be $\sim 100$ times the tidal force. 
Furthermore, consider a satellite before it has undergone tidal
stripping, so that its halo extends up to its original virial radius. 
In
this case it turns out that the ratio between the two forces is
$F_{\dot u}/F_{tidal}\sim 10 Yw(r_v)$. This result has far reaching
consequences as it shows that the LV force may provide a competing
mechanism for halo disruption, in particular if the timescale for
particle extraction due to the LV force is shorter than that of tidal
stripping. This expectation is indeed confirmed by numerical
simulations, see below.

\subsection{Numerical results}
\label{sec:numerical_results}


\begin{table}[!]
\begin{center}
 \begin{tabular}{l c c c c  c}
 \multicolumn{6}{c}{{\bf Parameters values}} \\
 \hline\hline\\
 & $Y$ &~& $r_{sc}$ [kpc] &~& $\kappa_A$  \\
 \hline\\
{\bf Satellites: }& $0.05 - 0.95$                &~~~& $1 - 35 $ &~~~& $10^{-13} - 10^{-9}$\\
 & &&&   &\\
{\bf Cluster: }& $0.05 - 0.95$                && $10 - 350 $ && $10^{-11} - 10^{-7}$\\
 & &   &\\
   \hline\hline\\
\end{tabular}
\end{center} 
\caption{Range of the LV parameters explored. $Y$ is the coupling
  strength between DM and the LV vector while $r_{sc}$ is the
  screening scale radius. The third column is the corresponding range
  for the parameter  $\kappa_A$. These tiny $\kappa$ values  
 are not constrained by any other observation.}\label{table:param}
\end{table}


\begin{table}
\begin{center}
 \begin{tabular}{c c l c}
  \multicolumn{4}{c}{Parameters of Draco DM halo} \\
 \hline\hline\\
  $V_t$ & 210 $\kms$ & tangential velocity of Draco satellite& \cite{Pryor:2014yoa}\\
 $\sigma_v$ & 10 $\kms$& velocity dispersion of Draco & \cite{Munoz:2005be,Walker:2007ju}\\
 $D$ & 82 kpc & distance of Draco from the galactic center & \cite{Mateo:1998wg} \\
 $T$ & $2.2$ Gyr & orbital period & \\
 $M_v$ & $6.2 \times 10^{8} M_\odot$ & Draco DM halo virial mass &   \\
  $r_s$ & 1.75 kpc & Draco actual DM halo radius [A] &  \cite{Moore:2001vq}\\
   $r_s$ & 17.5 kpc & Draco DM halo virial radius [B] & \\
 & &  & \\
   \hline\hline\\
    \multicolumn{4}{c}{Parameters of Fornax DM halo} \\
 \hline\hline\\
  $V_t$ & 220 $\kms$ & tangential velocity of Fornax satellite& \cite{Piatek:2002bb}\\

 $\sigma_v$ & 10 $\kms$& velocity dispersion of Fornax & \cite{Walker:2005nt,Walker:2007ju}\\
 $D$ & 147 kpc & distance of Fornax from the galactic center & \cite{Mateo:1998wg} \\
 $T$ & $4.3$ Gyr & orbital period & \\
 $M_v$ & $6.2 \times 10^{8} M_\odot$ & Fornax DM halo virial mass & \\
  $r_s$ & 1.75 kpc & Fornax actual DM halo radius & \cite{Moore:2001vq} \\
 & &  & \\
   \hline\hline\\
    \multicolumn{4}{c}{Parameters of a DM galactic halo in Coma cluster} \\
 \hline\hline\\
  $V_t$ & 1000 $\kms$ & tangential velocity of member galaxy& \cite{0067-0049-125-1-35,Lokas:2003ks}\\

 $\sigma_v$ & 120 $\kms$& velocity dispersion of galaxy DM particles &\\
 $D$& $2$ Mpc & distance of the galaxy from the Coma center & \\
  $T$ & $12$ Gyr & orbital period & \\
 $M_v$& $6.1\times 10^{11} M_\odot$ & galaxy virial mass &\\
 $r_s$ & 175 kpc & galaxy DM halo virial radius &  \\
 & &  & \\
   \hline\hline\\

\end{tabular}
\end{center} 
\caption{Halo parameters used in the integration of the equations of
  motion. See also \cite{McConnachie:2012vd} and references
  therein. } \label{tab:galaxy_data} 
\end{table}


In this section we present  results from the numerical integration of equation \eqref{eq:sat_p} of a DM halo orbiting inside the halo of its host galaxy. We consider a system similar to the MW and two of its dwarf satellites, Draco and Fornax. 
This choice is particularly suited for our purposes.  Draco is placed
at a galactocentric distance of $82$ kpc while Fornax is at $147$
kpc. The motion of Fornax  is compatible with a nearly circular orbit
\cite{Piatek:2002bb}. Draco motion is instead more controversial:
observations form the \textit{Hubble Space Telescope} placed it on a
relatively circular orbit \cite{0004-637X-790-1-74,Pryor:2014yoa}, but
more recent observations from \textit{Subaru} telescope suggest an
elliptical orbit 
\cite{2016MNRAS.461..271C}. Although this is an important aspect that
must be taken into account in a
more detailed study, it is not crucial for our analysis, so we assume
a circular orbit for Draco as well. In order to
make the 
investigation more complete and to explore a broader range of scales
and velocities, we also  consider the case of a test galaxy belonging
to the Coma cluster.  In Table \ref{table:param} we show the ranges of
parameters that these systems may be sensitive to. In particular,
assuming a certain range for the values of the screening length
$r_{sc}$ and of $Y$ determines through equation \eqref{eq:def_rsc}
the values of the LV parameter
$\kappa_A$ that one can explore. The range of parameters accessible
to our study is well
beyond those probed by other tests mentioned in section
\ref{sec:LVDM}. 

Table \ref{tab:galaxy_data} lists the parameters of
the halos that
have been used in the numerical integration.   
For Draco we explore  two scenarios with the same mass. Draco A:  all
the halo mass lies inside a radius of $\sim 2$ kpc which is the cutoff
radius seen in the observations of the stellar component
\cite{Moore:2001vq}. Draco B:  the halo extends to a radius of $\sim
20$ kpc which corresponds to the virial radius of a halo with velocity
dispersion of $\sim 10\kms$ consistent with the observations of
Draco. These two scenarios may be seen as  corresponding to
satellites with and without gravitational tidal stripping having
occurred as they settle in their  final orbit in the MW. In the case
of Fornax the radius of the DM halo is the same as that of Draco
A. The table also gives the parameters we adopt for the system of  a
galaxy in a cluster like Coma  ($R_\text{Coma}\sim 3$ Mpc)
\cite{Lokas:2003ks}. In all cases the virial radii reported in table
\ref{tab:galaxy_data} are consistent with the  observed one
dimensional velocity dispersion.

\begin{figure}[!t]
 \includegraphics[scale=.8]{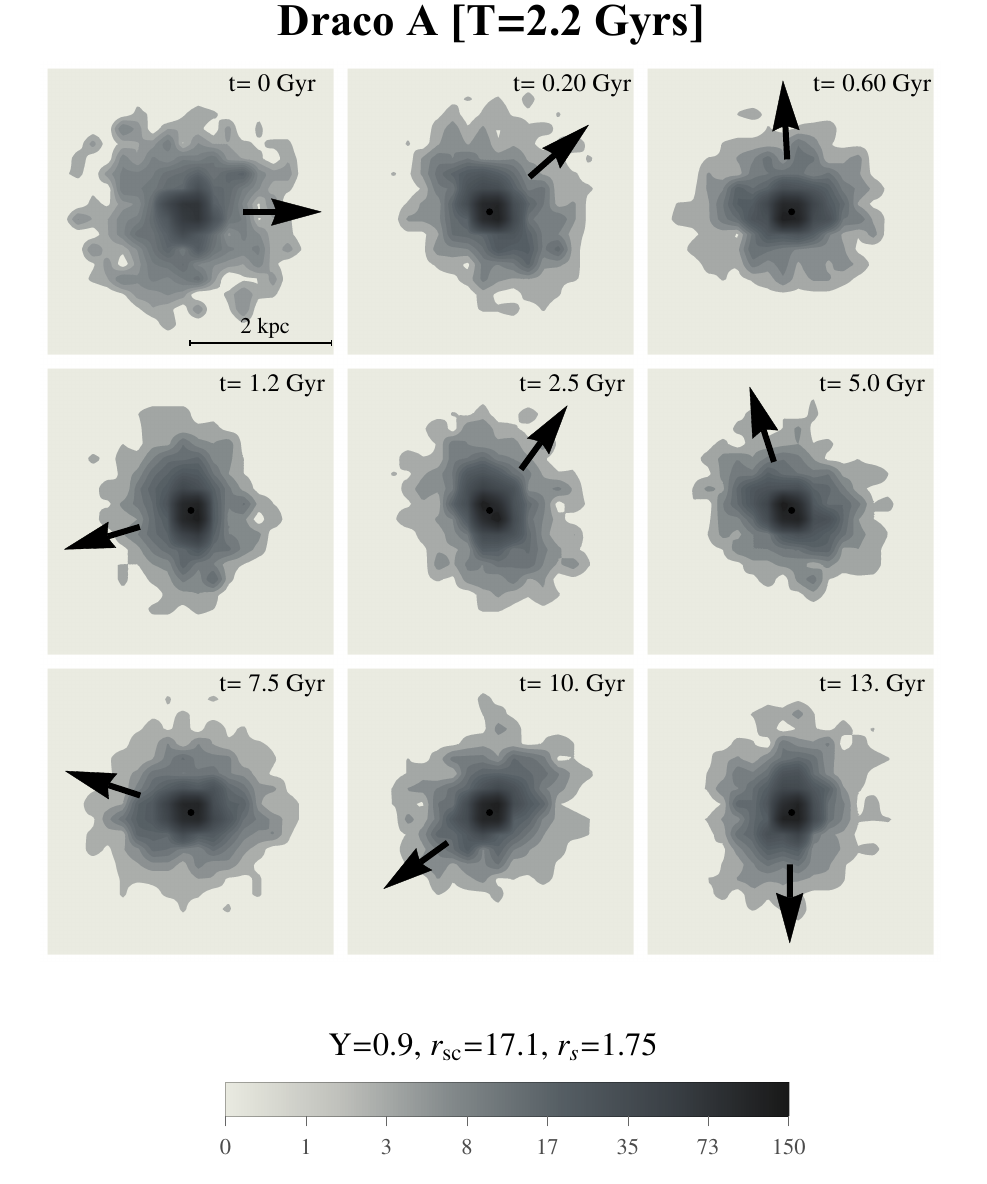} \hfill
   \includegraphics[scale=0.8]{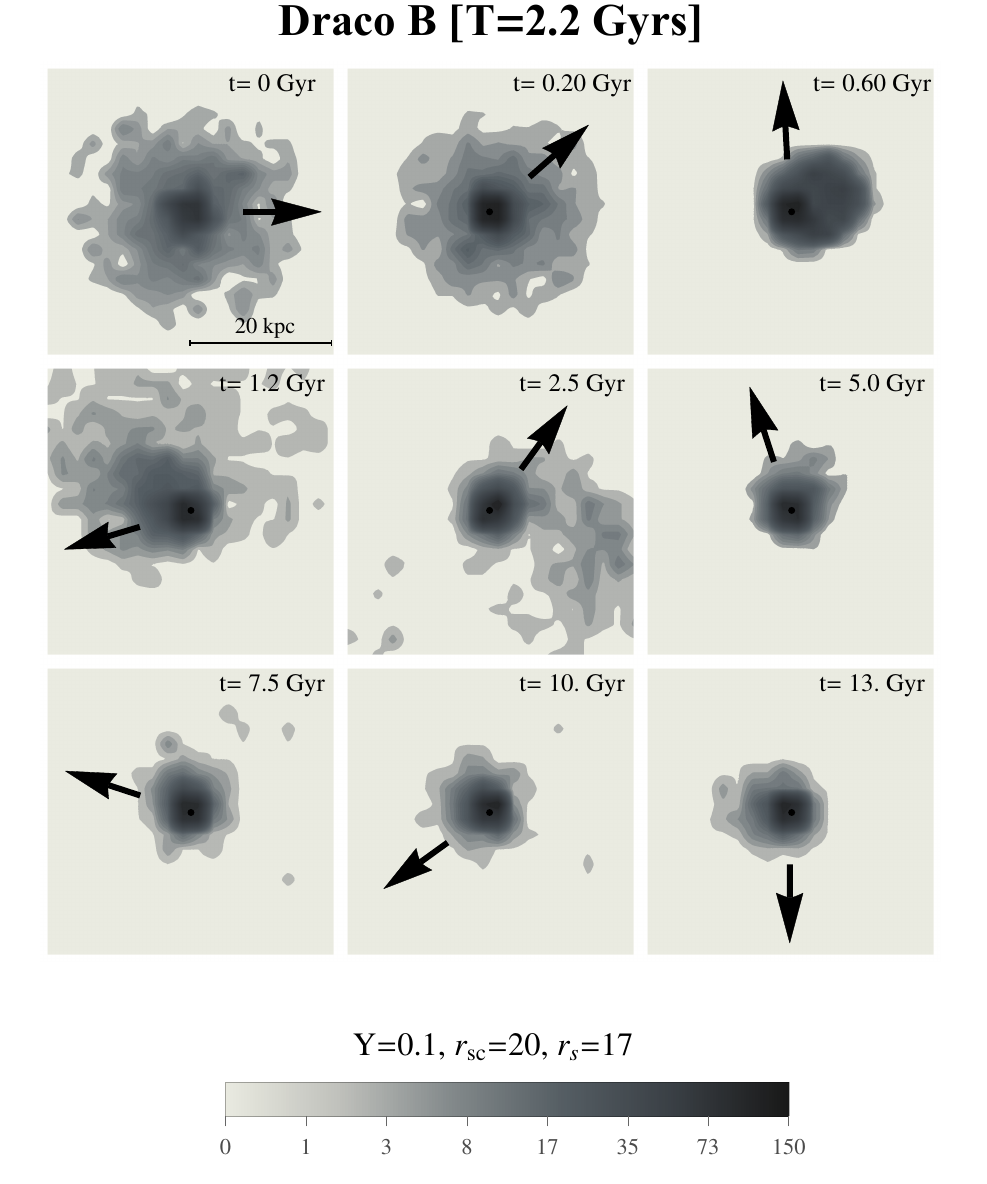} \\
  \includegraphics[scale=.8]{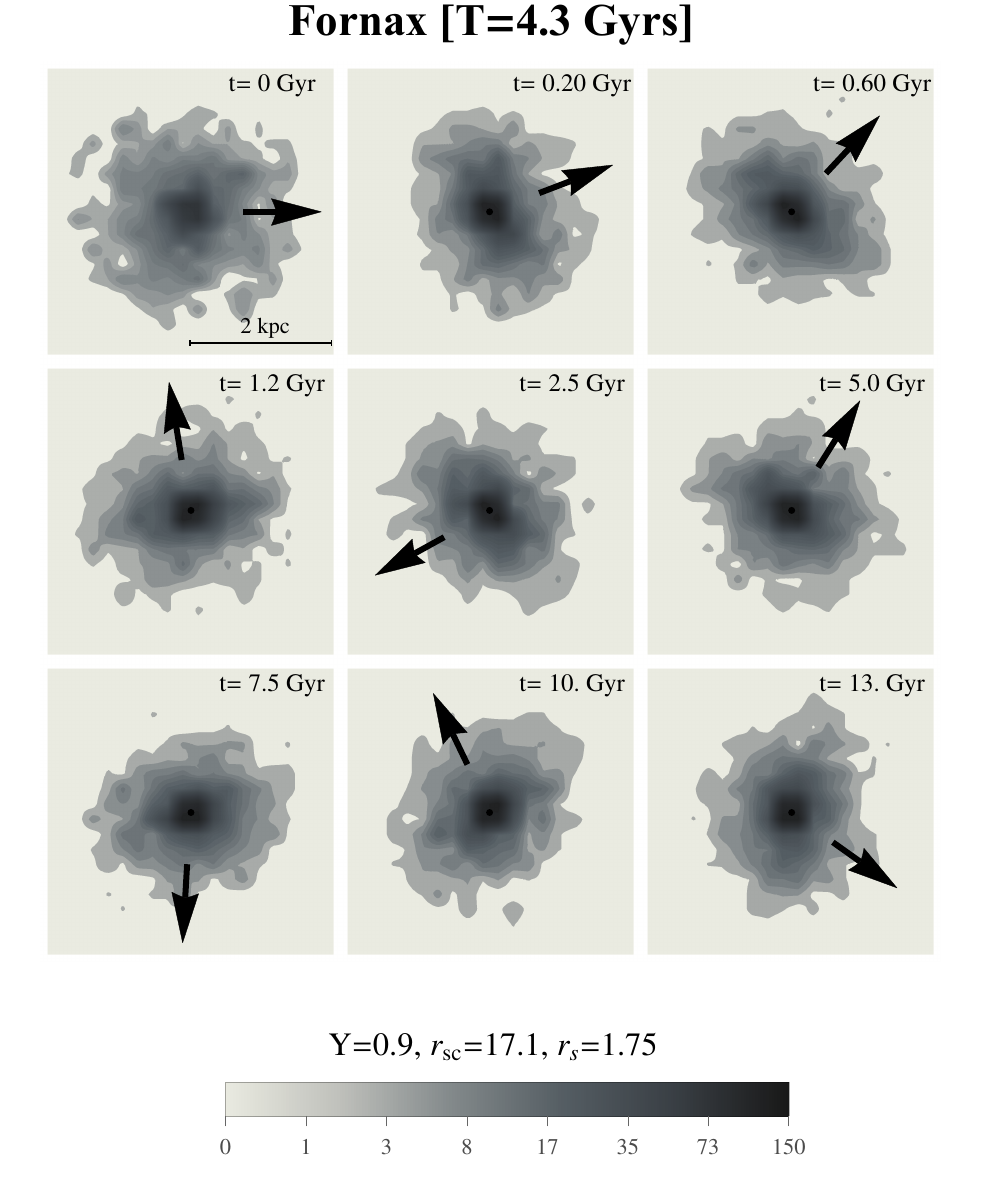} \hfill
   \includegraphics[scale=.8]{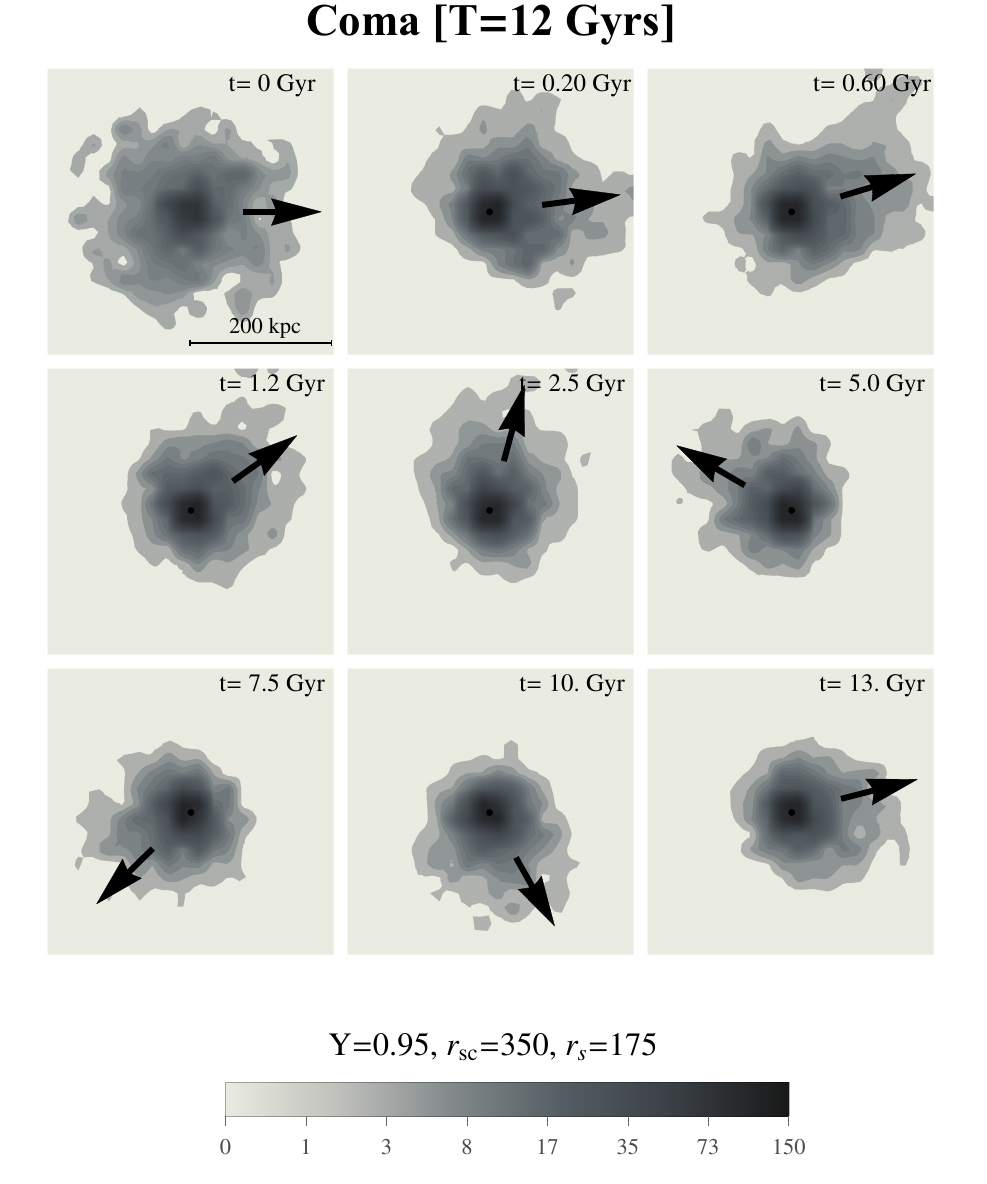}
 \caption{Particle distribution in the orbital plane of the satellite
   from the simulation output at different times. The 4 subplots
   correspond to the 4 considered systems with orbital periods $T$, 
as indicated in the figure.
 The  top left panel in each subplot represents the initial
 configuration. The black arrow  points towards the center of the host
 halo.
} \label{fig:nocore_xy}
\end{figure}

The numerical solution is obtained by means of a 4(5)~Runge--Kutta
integration method over a Hubble time.  
We have considered both the gravitational potential and the LV force
as external functions  assuming that backreaction is negligible. This
is a good approximations as long as departures from spherical symmetry
are moderate.

To simulate the satellite halo, we have used 1500 particles with
randomly oriented initial velocities. Their initial positions are
adjusted in such a way that, under the influence of gravity alone, 
they would stay on circular orbits around the center of the
satellite. The initial density distribution of particles follows 
an Einasto profile
\cite{1965TrAlm...5...87E,RetanaMontenegro:2012rq} for which the mass
enclosed inside a  radius $r$ is 
\begin{equation}
\label{eq:einasto}
M(r)=\left(1-\frac{\Gamma(3 n_E,(r/h)^{1/n_E})}{\Gamma(3 n_E)}\right)\left(1-\frac{\Gamma(3 n_E,(r_s/h)^{1/n_E})}{\Gamma(3 n_E)}\right)^{-1}\,,
\end{equation}
where $\Gamma $ is the incomplete gamma function, the mass is normalized such that  $M(r=r_s)=1$ and the   parameter $h$ satisfies  $h=r_{-2}/(2 n_E)^{n_E}$ where $r_{-2}$ is the radius at  which  the  slope $d\log \rho/d\log r=-2$. 
The slope parameter $n_E$ decreases with halo mass and is  $4.54
\lesssim n_E \lesssim 8.33$ \cite{Hayashi:2003sj}, where the lower and
higher values correspond to a cluster and a dwarf halo, respectively.
In our simulation we adopt $n_E = 5.88$ and $r_{-2}= r_v/20$.  

The LV force is provided by the potentials found in the previous
section, namely equations \eqref{eq:genv} and
\eqref{eq:genoutv}. Notice that these have been derived assuming the
particle distribution in the satellite follows an $r^{-2}$ profile
rather than an Einasto. However, in the situations we will consider,
the deviations of the Einasto profile from the inverse square law are
small and the two profiles are similar over the part of the satellite
containing most of the mass.

\subsubsection{Particle distribution}\label{sec:eff_part_dist}

We begin with a  general  visual assessment of   the particle
distribution. This will serve as an introduction to the more
quantitative analysis below.  

Figure \ref{fig:nocore_xy} shows the particle distribution in the
plane of 
motion of the satellite at different times.
The halo taken to be spherically symmetric at $t=0$ is deformed at later
times in a very specific way.  
The circular motion of the satellite produces a periodic LV force
which turns the halo into elliptical shape with a time varying
orientation. Another effect of the LV force is to enhance the particle
distribution density at the center. Interestingly, a stable
configuration is reached in all cases after 5 Gyr at the latest. 
 
For Draco B the effect of LV force is particularly dramatic, as can be
seen from figure \ref{fig:nocore_xy} (top right). At the beginning
  we see a transient phase due to the fact that particles are
  initially placed on stable orbits according to gravity
  alone. Quickly after that the 
LV force kicks in and rapidly extracts particles from the outer
shells, whereas the particle distribution in the central region gets
tighter. 
We emphasize that this effect is purely due to the LV interaction, as
the tidal gravitational force has been switched off in the simulations
(see the discussion preceding equation \eqref{eq:sat_p}).
Although
this is a very extreme case, it is nonetheless instructive since  it
shows how the LV force may provide a very efficient mechanism for disrupting
shallow sub-halos. 
This mechanism can operate on timescales shorter than the usual
tidal disruption and, unlike the latter, gives rise to 
asymmetric tails of debris. 
A  galactic halo in the Coma cluster  exhibits  similar: rapid mass
extraction from the outer shell  and then the halo settles to a stable
configuration with a smaller radius, figure ~\ref{fig:nocore_xy} 
(bottom right). 

These examples demonstrate several interesting consequences of the LV
force: it changes the density profile of the halo, it may remove a
significant amount of particles and it changes the initially spherical
distribution.  
In the next three subsections we will quantify these effects by analyzing
the mass distribution inside the halos, the mass loss and 
the ellipticity of the halos as functions of the model
parameters. 

\subsubsection{Radial mass profiles}
\label{sec:constr_mass_distrib}

 We examine now how the mass distribution of the satellite halo is affected  by the presence of the LV force. 
 We compute   the  mass  $M(r)$ within spherical shells from the  particle distribution  in the simulation. We select the simulation output  at a  sufficiently late time to ensure that the system has reached  a steady state. 
In the normalization of $M(r)$ we include only  particles inside the
outer radius of the initial configuration.

\begin{figure}[!ht]
         \includegraphics[scale=0.6]{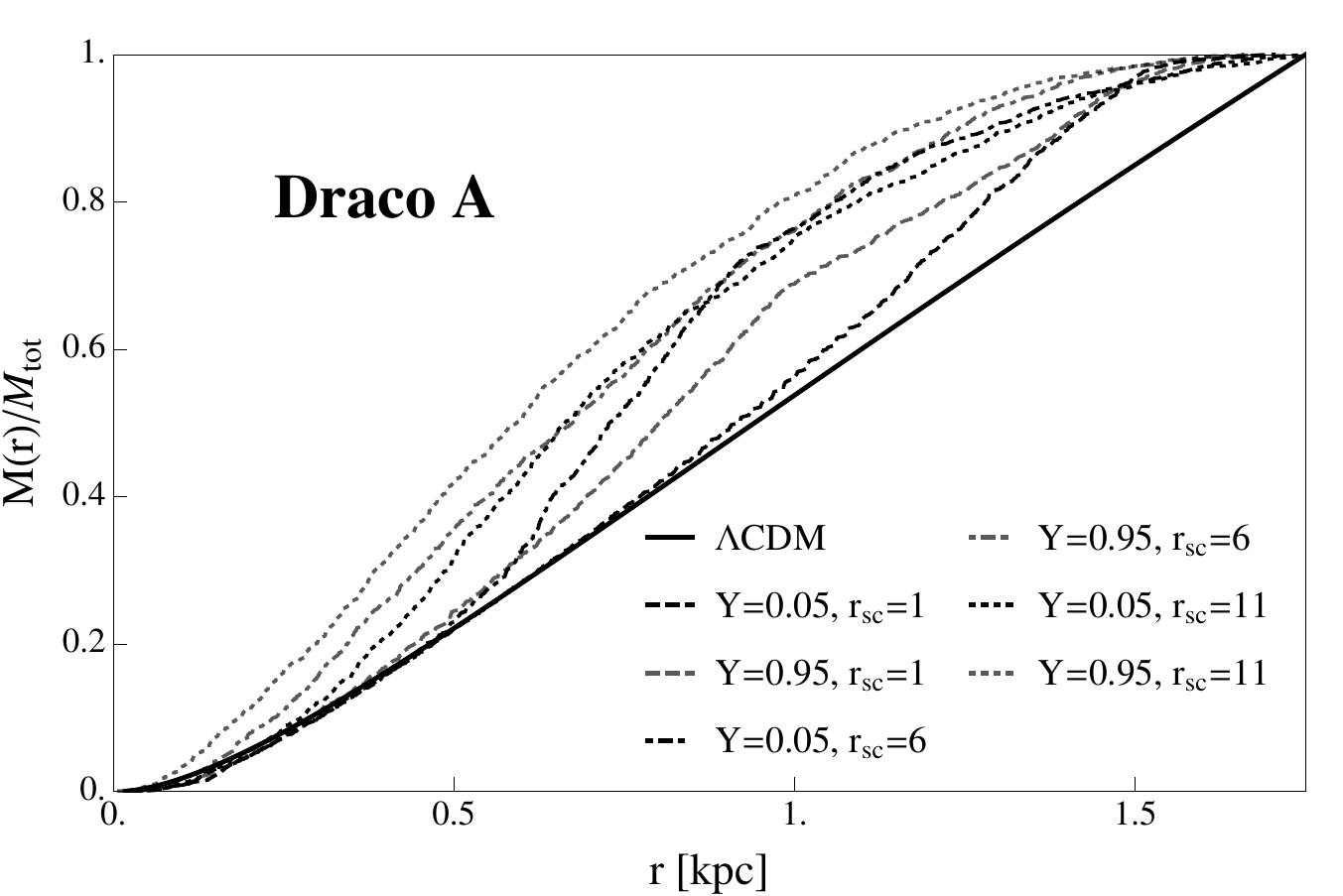}     
        \includegraphics[scale=0.6]{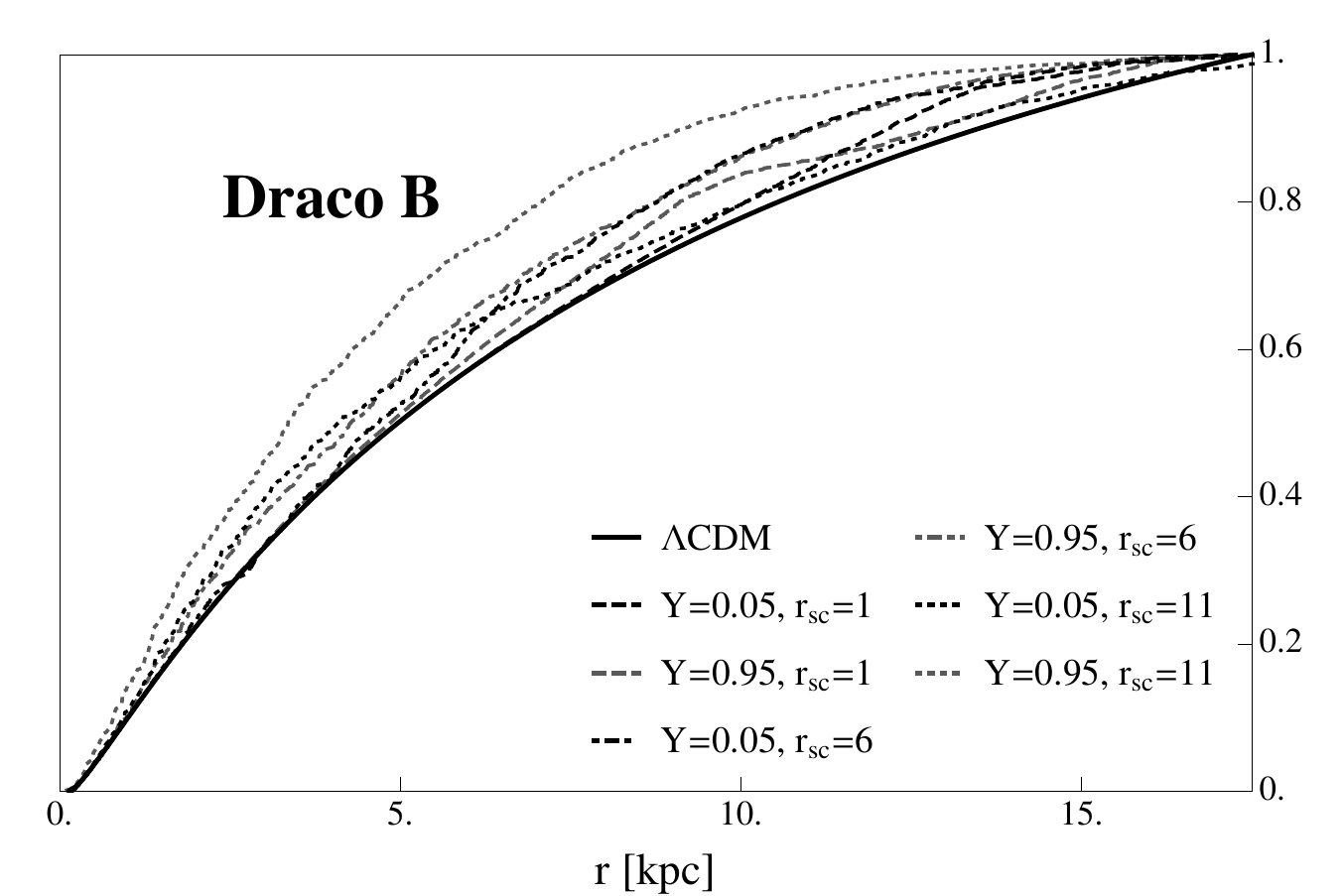} \\
         \includegraphics[scale=.6]{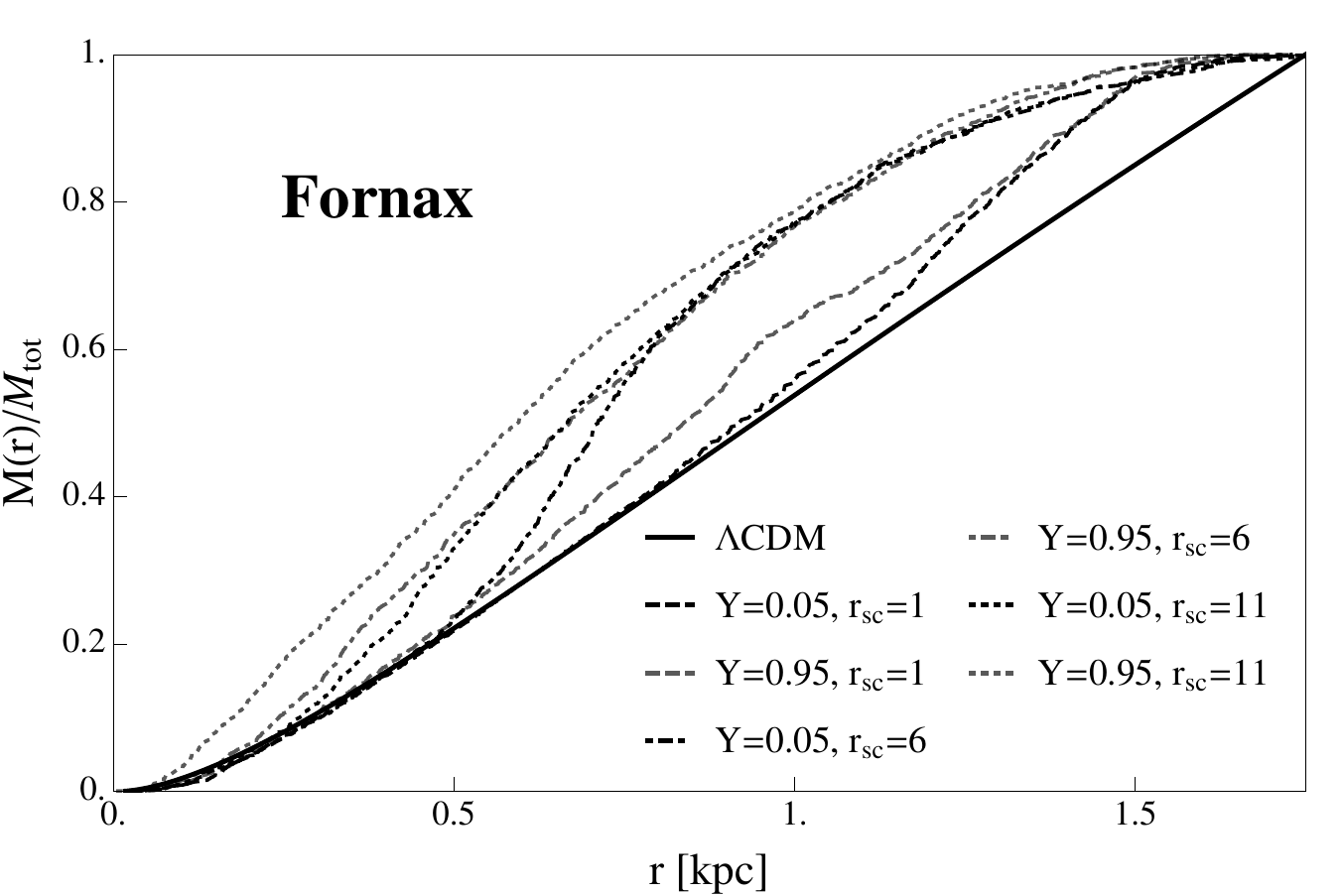} 
        \includegraphics[scale=.6]{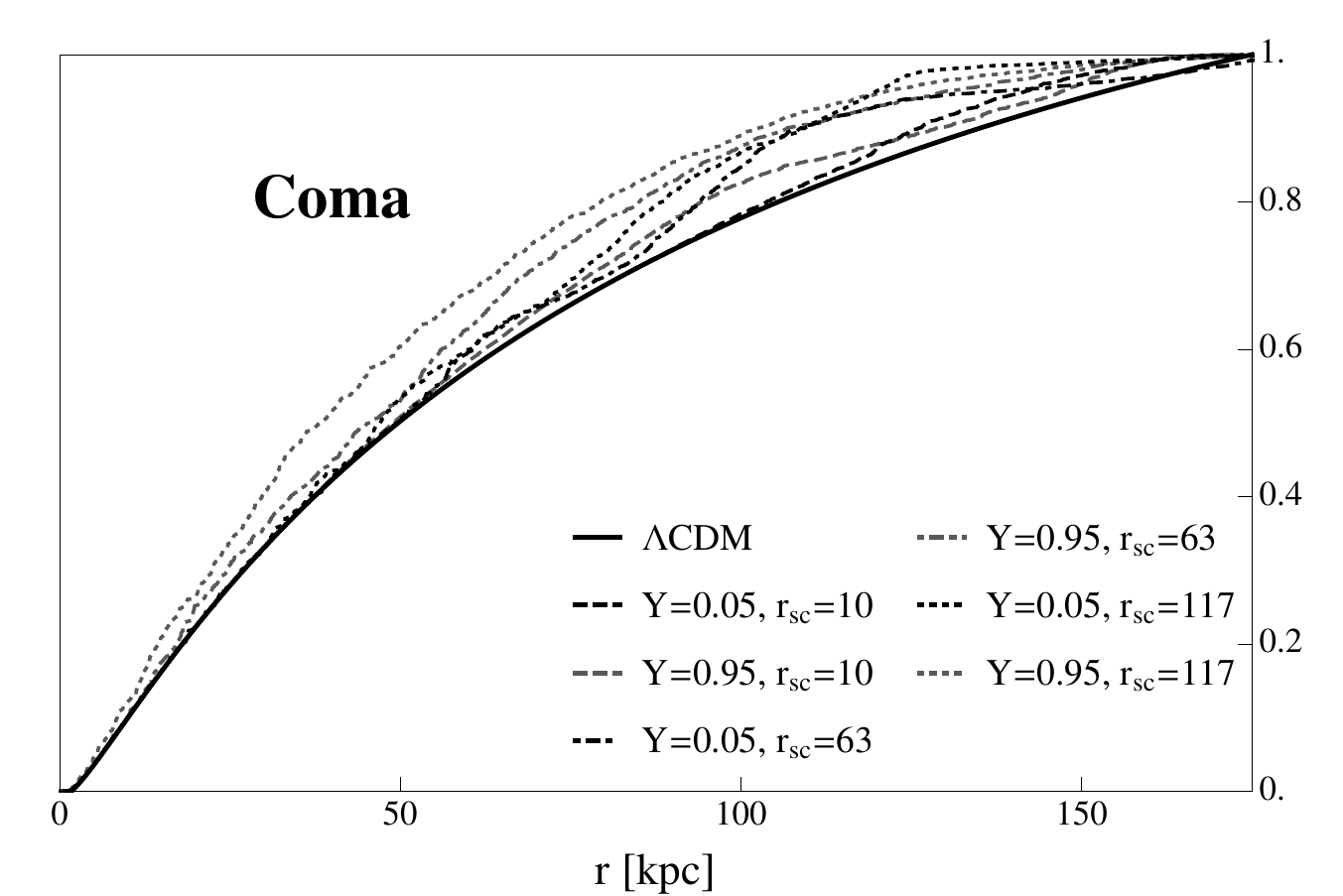} 
    \caption{Mass within a given radius for  Draco A (top left),
      Draco B (top right), Fornax (bottom left) and a galaxy in the
      Coma cluster (bottom right) for various values of $r_{sc}$ and
      $Y$. The black  line in each panel shows
 the initial distribution.} \label{fig:Draco_strip_nocore_Mr}
  \end{figure}

 Curves of $M(r)$ are plotted in figure
 \ref{fig:Draco_strip_nocore_Mr} for all systems. 
 The solid line in each  panel refers to the initial unperturbed
 profile. This profile would be maintained if the system were advanced
 according to gravity alone\footnote{This is true also in the case of
   Draco B because we are neglecting the effects due to the tidal
   forces of the host halo.}. 
The other curves correspond to  the
 particle distribution  evolved with the  LV force for several values
 of $r_{sc}$ and $Y$, as indicated. 
 The LV force introduces significant changes to the mass distribution
 of the subhalo.  
This confirms the visual assessment of 
Fig.~\ref{fig:nocore_xy}: 
the evolved satellites are denser and more compact with respect to the solid line. 
For a quantitative assessment of the particle distribution we fit the
evolved mass profile $M(r)$ by  an  
Einasto functional form (cf. \eqref{eq:einasto})  with both $n_E$ and
$r_{-2}$ as free parameters.  
We find that, apart from the central regions, the Einasto profile
provides a reasonable fit.  However, the best fit values of   $n_E$
and $r_{-2}$ deviate  dramatically  
from the observationally motivated range. 
For the values of  $r_{sc}$ and $Y$ considered in the figure, the fit
yields $0.61 \lesssim n_E \lesssim 1.62$ and $0.36 \lesssim
r_{-2}\lesssim 0.71$ for Draco A, $2.4 \lesssim n_E \lesssim 4.94$ and
$0.82\lesssim r_{-2}\lesssim 2.02$ for Draco B, $0.61 \lesssim n_E
\lesssim 1.9$ and $0.33 \lesssim r_{-2} \lesssim 0.71$ for Fornax and
$2.35 \lesssim n_E \lesssim 4.41$ and $12.82 \lesssim r_{-2}\lesssim
20.83$ for a galaxy in Coma cluster. 
\begin{figure}[!ht]
\centering
\includegraphics[scale=1.]{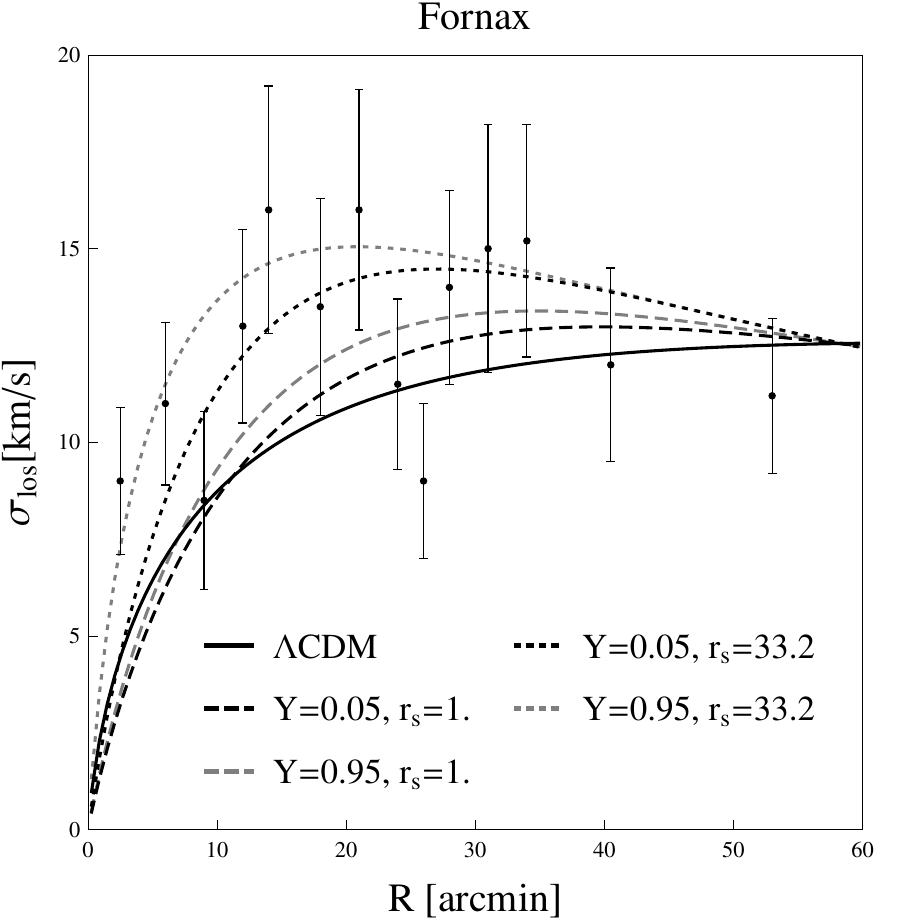}
\caption{Theoretical predictions for the line-of-sight velocity dispersion
  compared with the observed values. The continuous line corresponds to
  $\sigma_{los}$ obtained from an Einasto profile with $n_E=5.88$
  and $r_{-2}=r_v/20$ in $\Lambda$CDM model. The dashed and dotted
  lines correspond to different choices for the LV parameters. The
  data points and the $1\sigma$ error bars have been taken from
  \cite{Mateo:1997rs}.} 
\label{fig:sigmalos_Fornax}
\end{figure}

Caution should  be exercised  in interpreting these results as they
have been derived in a simplified model, which does not take into
account the formation process of the satellite. 
It is nonetheless interesting
to see if one can constrain the LV effect from observations of the
stellar component. In particular,  
 good measurements of the line-of-sight velocity dispersion of stars
 in  Fornax are  
available  \cite{Mateo:1991xx,Mateo:1997rs,Walker:2005nt}. Even if
stars are not influenced by the LV force, their observed distribution
and velocity dispersion  
are a probe of the underlying DM distribution through their
gravitational interaction.  
Given the DM distribution in the simulation, one can thus predict 
the velocity dispersion of baryonic tracers (stars). 
Assuming circular orbits for the stars, we computed their line of
sight velocity dispersion  according to
\cite{1982MNRAS.200..361B,1987gady.book.....B,Lokas:2001gy} 
\begin{equation}
\sigma^2_{los}(R)=\frac{1}{I(R)}\int_{R}^{\infty} \left(\frac{R}{r}\right)^2 \frac{\nu(r)\, g(r) r^2}{\sqrt{r^2-R^2}}\,dr\,,
\end{equation}
where $R$ is the projected distance from the subhalo center,
$I(R)$ is the surface brightness \cite{Sersic:1968xx,Ciotti:1991xx},
and  $\nu(r)$ is the 3D density of stars \cite{Neto:1999gx}. The
gravitational acceleration   $g(r) $  is due to the gravity of  
the DM in  the simulation and a subdominant  stellar component whose
contribution is estimated using the  observed  stellar distribution in
Fornax  \cite{Irwin:1995tb}.  
Figure \ref{fig:sigmalos_Fornax} shows  $\sigma_{los}(r)$  for $M(r) $
corresponding to the initial Einasto profile as well as evolved
particle distributions with several values of the LV parameters.  
The data shown in the figure do not allow to
constrain the parameter space of the LV
force. By looking at the plot one
may be tempted to conclude that some of the LV curves fit  the data
better than the unperturbed one. However, to draw
robust conclusions one should use more recent and precise data  
\cite{Walker:2009zp} which, in turn, require more precise
simulations
and a  broader investigation of the
relation between velocity dispersion and  
the underlying mass profile.  

\begin{figure}[!ht]
         \includegraphics[scale=0.7]{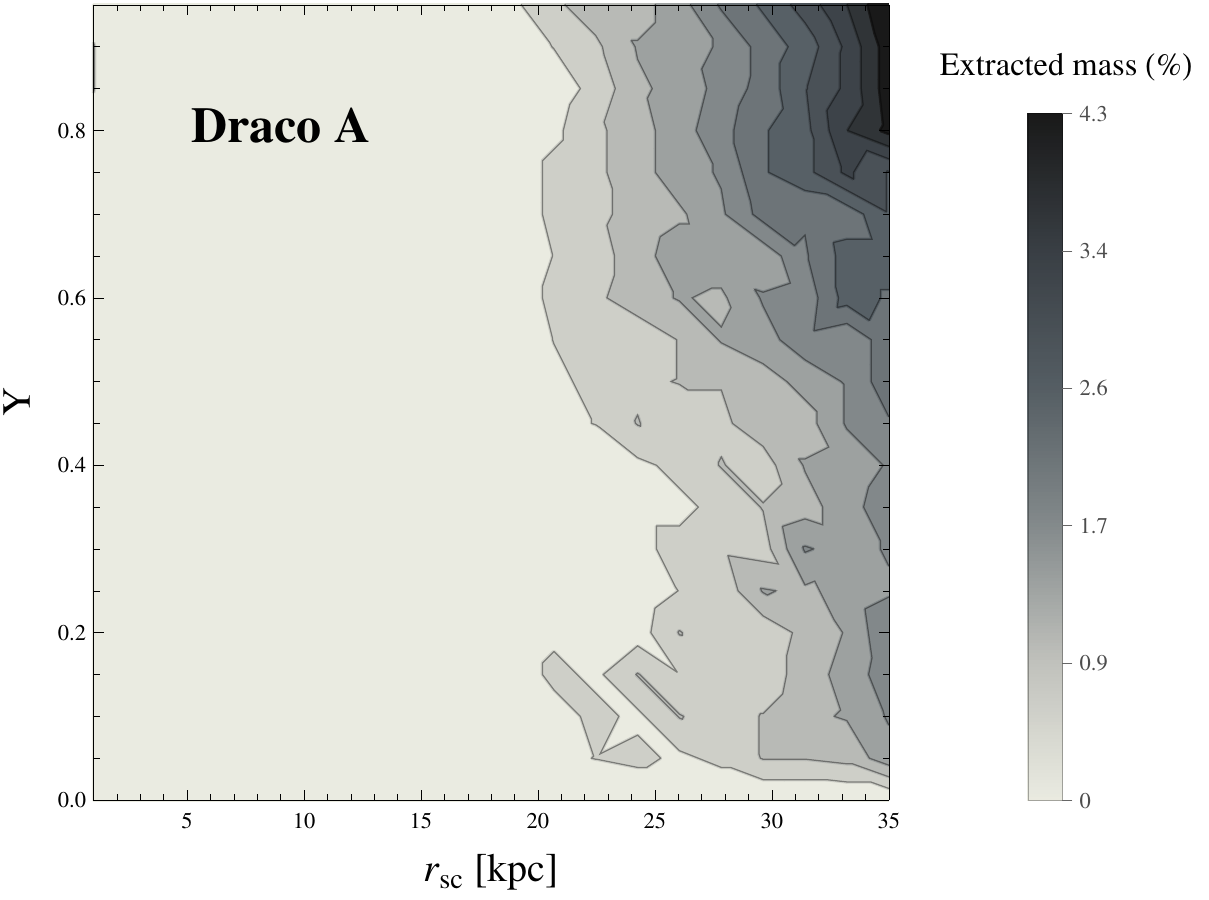}      \hfill
        \includegraphics[scale=0.7]{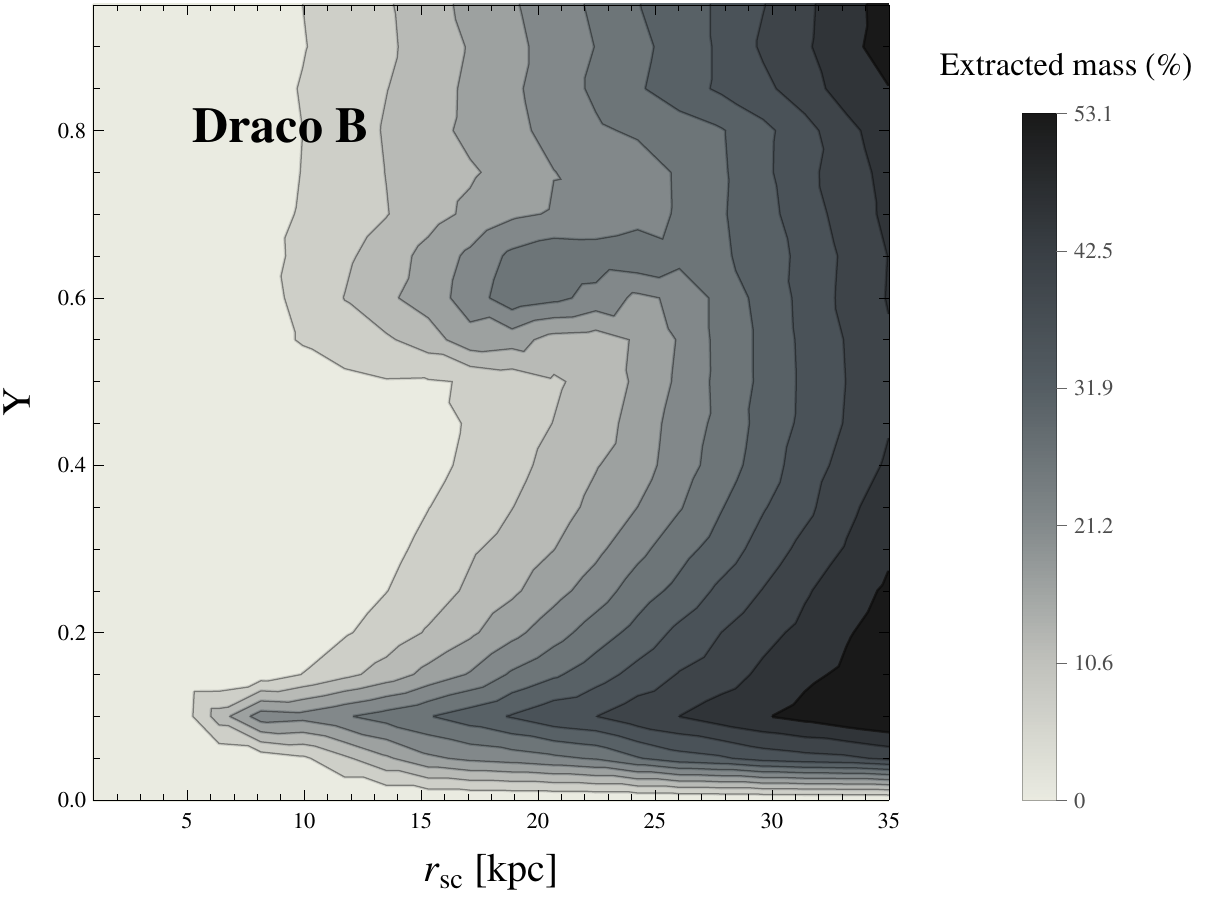} \\
         \includegraphics[scale=.7]{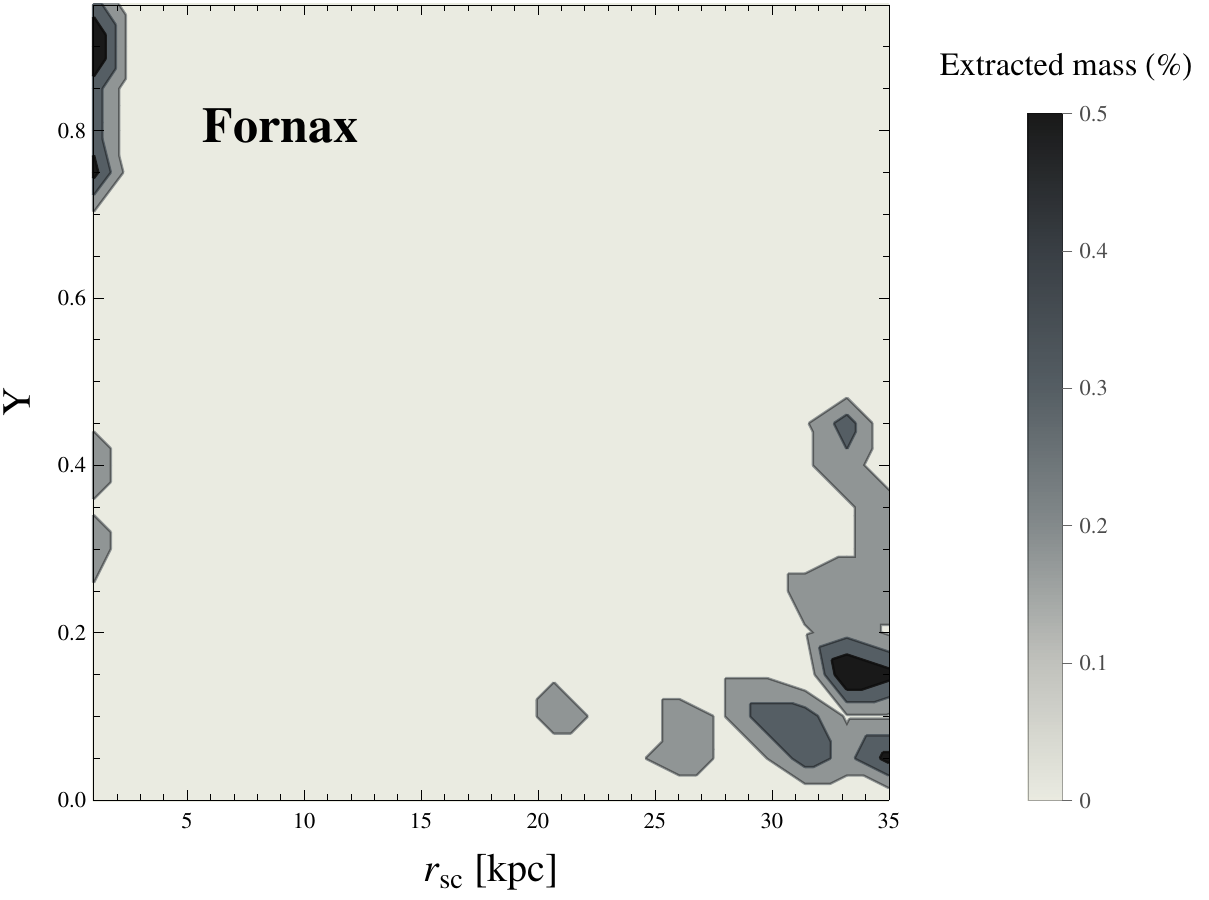} \hfill
        \includegraphics[scale=0.7]{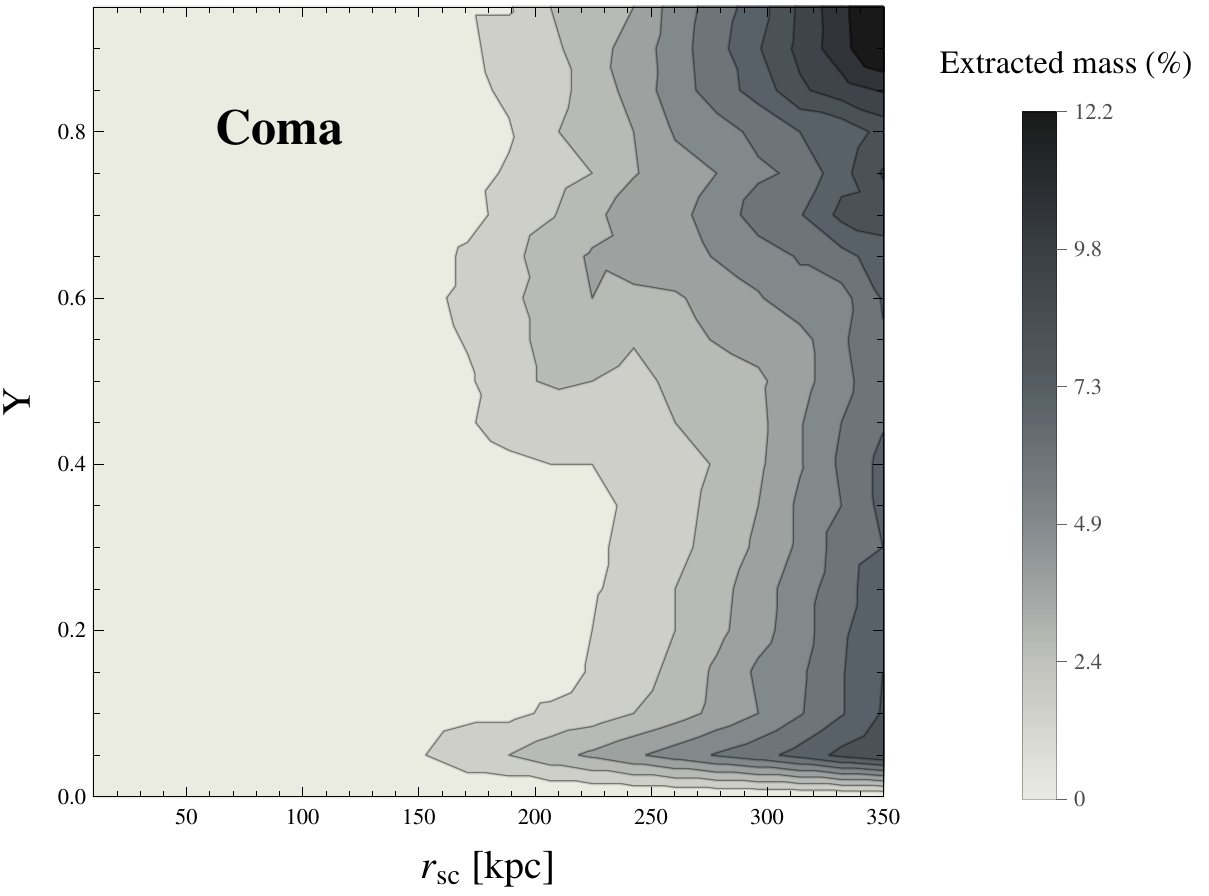} 
    \caption{Extracted mass for the stripped Draco halo (top left),
      the unstripped Draco halo (top right), Fornax halo (bottom left)
      and a galaxy in the Coma cluster (bottom right) for various
      values of the parameters $r_{sc}$ and
      $Y$.} \label{fig:extracted_mass} 
  \end{figure}

\subsubsection{Mass extraction}

We now quantify the mass extracted from the halo by the LV force as
the mass that lies outside the initial radius of the object at a given
time\footnote{Notice that this does not imply that all particles
  carrying this mass are not bound to the object anymore. However, we
  have checked that there is no significant change in the enclosed
  mass after a few time steps.}. Figures \ref{fig:extracted_mass} plot
the mass loss after a Hubble time as a function of the parameters
$r_{sc}$ and $Y$.  As can be seen from the plots, for the most compact
satellites the mass loss is small, almost non existing in the case of
Fornax. However, it is very pronounced for extended subhalos, reaching
a peak of $50\%$ in Draco B. This last case is particularly interesting because most of the mass is extracted during the first orbit, cf. Fig.~\ref{fig:nocore_xy}. In comparison, the amount of mass
extracted by tidal stripping is as high as $60\%$ of the initial mass
after the first pericentric passage \cite{Mayer:2007xb}. 
In general, the typical timescale of mass extraction by the LV force will depend on the particular set of parameters but we found that in many cases it is shorter than the one associated to tidal effects.  Another interesting effect is the dependence on the distance from
the center of the host galaxy.  Even if Draco A and Fornax have a similar initial configuration, the former experiences
a larger mass loss since it is closer to the center of the MW.

\subsubsection{Ellipticity}

In  \ref{sec:eff_part_dist} we have mentioned 
deviations from a spherical shape as 
another observable signature of LV. 
In order to assess these deviations  quantitatively, we have computed the ellipticity parameters from  the particle distribution in the simulation output.
We define a coordinate system ($x,y,z)$
with origin centered at and comoving  with the core of the
satellite. The directions of the 3 axes are  fixed with respect to the
host halo. Furthermore, the $xy$ plane is defined by the orbital
motion of the satellite inside the host halo. 

The ellipticity is characterized by the quadratic moments of the
density distribution,
\begin{equation}
Q_{ij}=\frac{1}{N}\sum_{p=1}^N x^i_p\,x^j_p\,,
\end{equation}
where the summation is over all $N$  particles of the subhalo.
We define the ellipticity parameters in the plane ($x,y$) 
as
\begin{equation}
\label{eq:Qij}
(\epsilon_1,\epsilon_2)=\left(\frac{Q_{xx}-Q_{yy}}{Q_{xx}+Q_{yy}+2(Q_{xx}Q_{yy}-Q_{xy}^2)^{1/2}}\,,\;\frac{2Q_{xy}}{Q_{xx}+Q_{yy}+2(Q_{xx}Q_{yy}-Q_{xy}^2)^{1/2}}\right)\, ,
\end{equation}
and similarly for the plane ($x,z$). 
Denoting the major and minor axes by $a$ and $b$ respectively, the
ellipticity parameters can be written  as 
\begin{eqnarray}
\epsilon_1 & = & \frac{1-q}{1+q}\cos(2\theta)\,,\\
\epsilon_2 & = & \frac{1-q}{1+q}\sin(2\theta)\,,
\end{eqnarray}
where  $q=b/a$,  and  $\theta$ is the angle between the major axis and
the 
$x-$direction.

\begin{figure}[!ht]
\includegraphics[scale=0.5]{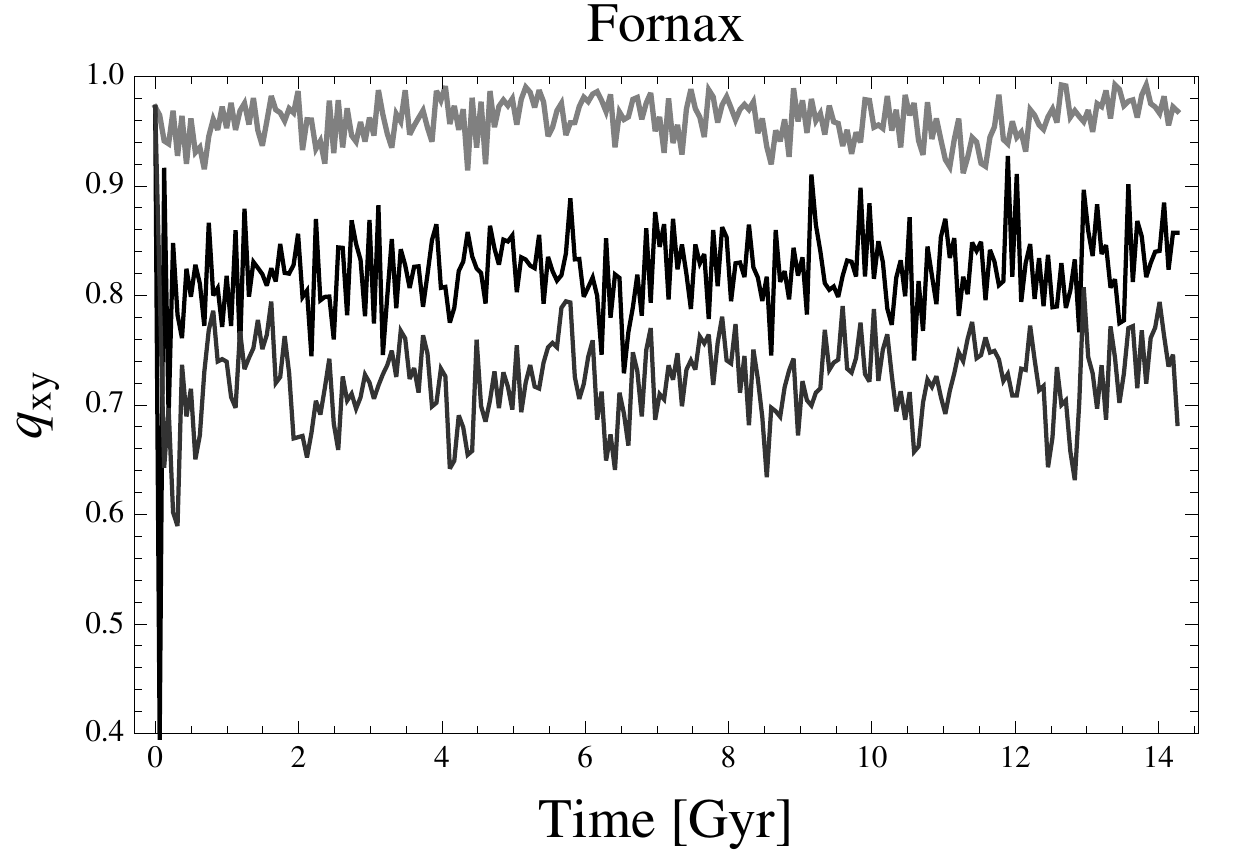}\hfill
\includegraphics[scale=.6,trim=0cm -1.5cm 0cm 0cm]{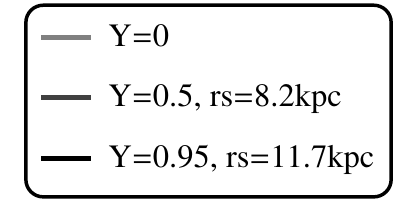}\hfill
\includegraphics[scale=0.5]{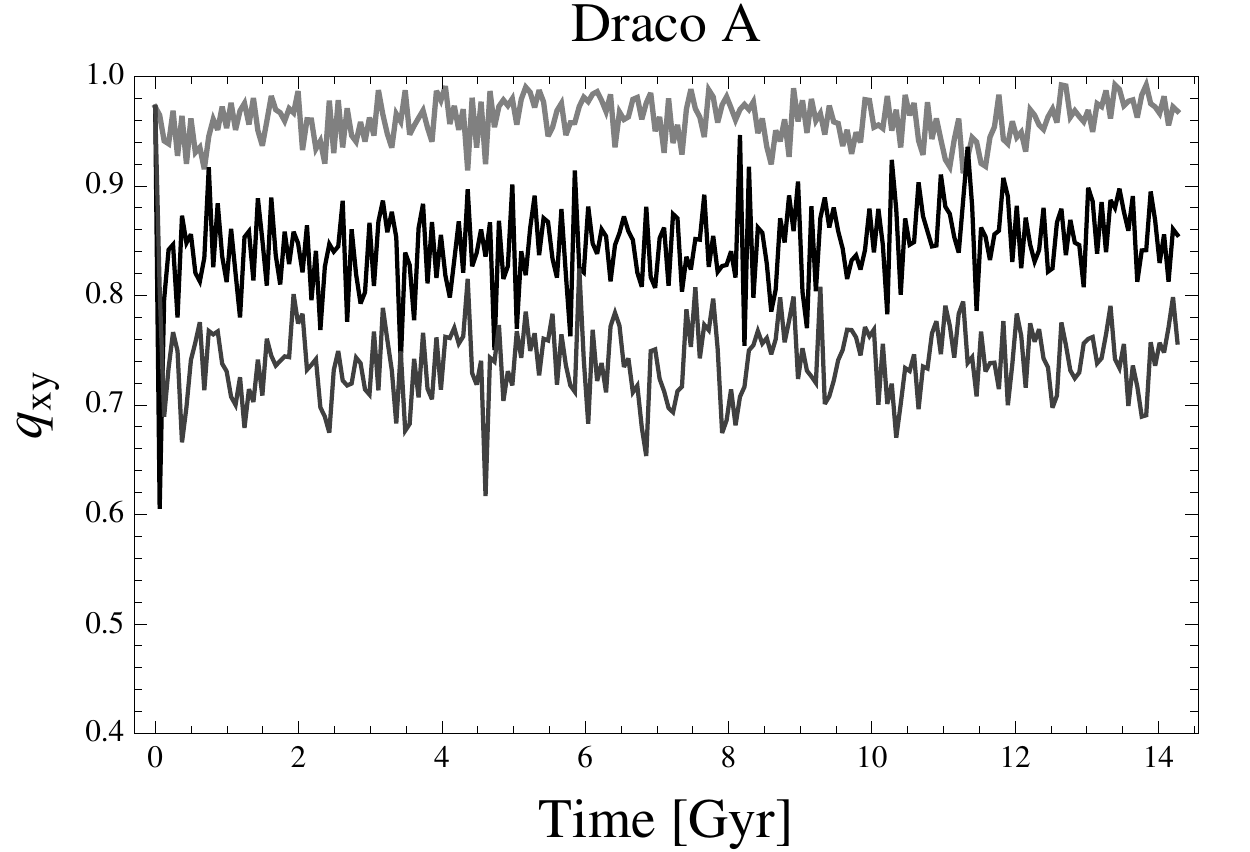}
\caption{Time evolution of  $q=b/a$ parallel to the orbital plane 
 for three sets of values of the parameters $(Y,r_{sc})$: $(0,-)$ light gray, $(0.5,8.2 \,\text{kpc})$ dark gray and $(0.95,11.7 \,\text{kpc})$ black line for  Fornax (left) and Draco A (right).}
 \label{fig:qxy_vs_time}
\end{figure}

\begin{figure}[!ht]
\includegraphics[scale=0.5]{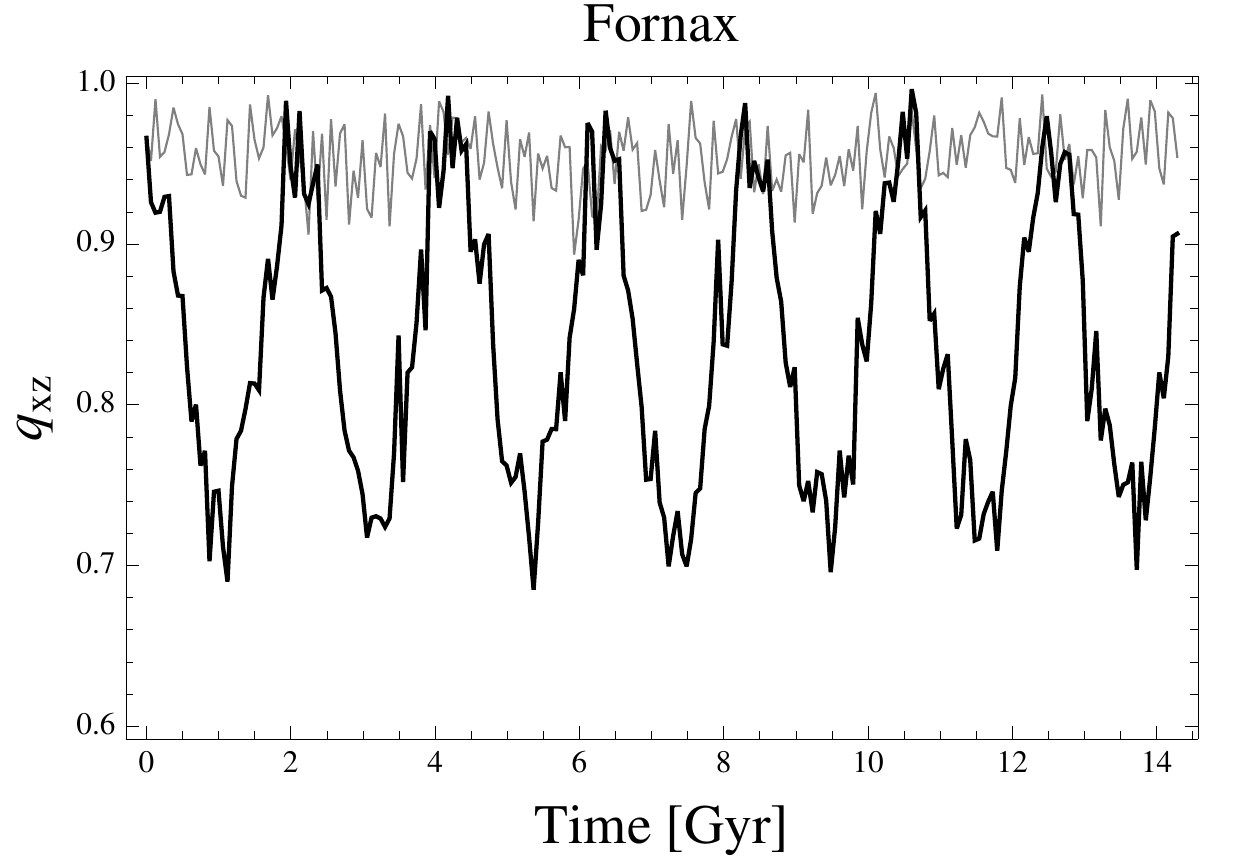}\hfill
\includegraphics[scale=.7,trim=0cm -1.5cm 0cm 0cm]{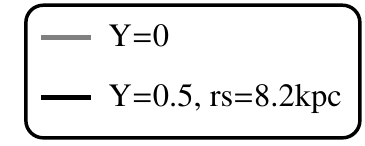}\hfill
\includegraphics[scale=0.5]{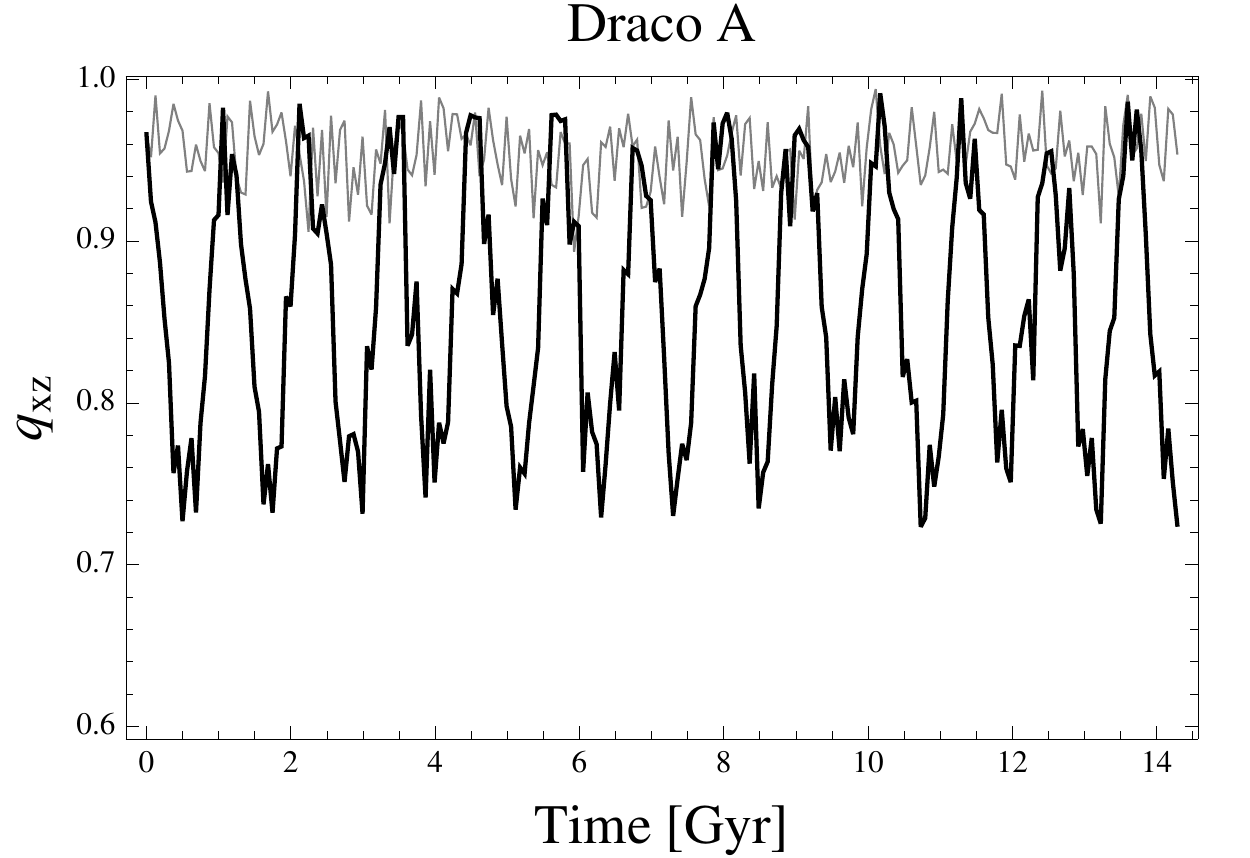}
\caption{Time evolution of  $q=b/a$ in the plane orthogonal to the orbital motion 
 for two sets of values of the parameters $(Y,r_{sc})$: $(0,-)$ light gray, $(0.5,8.2 \,\text{kpc})$ black line for  Fornax (left) and Draco A (right).}
 \label{fig:qxz_vs_time}
\end{figure}

We have computed the ellipticity parameters for Fornax and Draco for
$Y=0.95$,
$r_{sc}=11.7$~kpc and $Y=0.5$, $r_{sc}=8.2$~kpc.
In figure \ref{fig:qxy_vs_time} we plot the time evolution of $q$ in
the orbital plane of the satellite. 
The irregularities in the curves are due to the discreteness of the
particle distribution. 
The light gray curve in the plots corresponds to no LV force i.e. $q=1$.
The other curves reach values 
 substantially different from unity, almost immediately after the
 force is switched on and stay constant over a whole Hubble time.

On the other hand, the ellipticity parameter $q$ in the orthogonal
plane $(x,z)$ exhibits pronounced oscillations at approximately half the
orbital period, see Fig.~\ref{fig:qxz_vs_time}. Recall that the plane
$(x,z)$ is fixed with respect to the host halo. Then the dependences
shown in Figs.~\ref{fig:qxy_vs_time}, \ref{fig:qxz_vs_time} imply that
the satellite has a prolate ellipsoidal shape with the major axis
lying in the orbital plane and sustaining an almost constant angle
with the direction of motion.

The values of ellipticity obtained in our simulations
are compatible with  
observations
 \cite{2008ApJ...684.1075M,ellipticity}. 
Nevertheless they can be also explained within pure $\Lambda$CDM  
\cite{2014MNRAS.439.2863V,2015MNRAS.447.1112B}.
A more refined analysis is needed in order to tell
whether the ellipticity generated by the LV force exceeds the one
produced by gravitational forces alone.

\begin{figure}[!ht]
\centering
\includegraphics[scale=0.7]{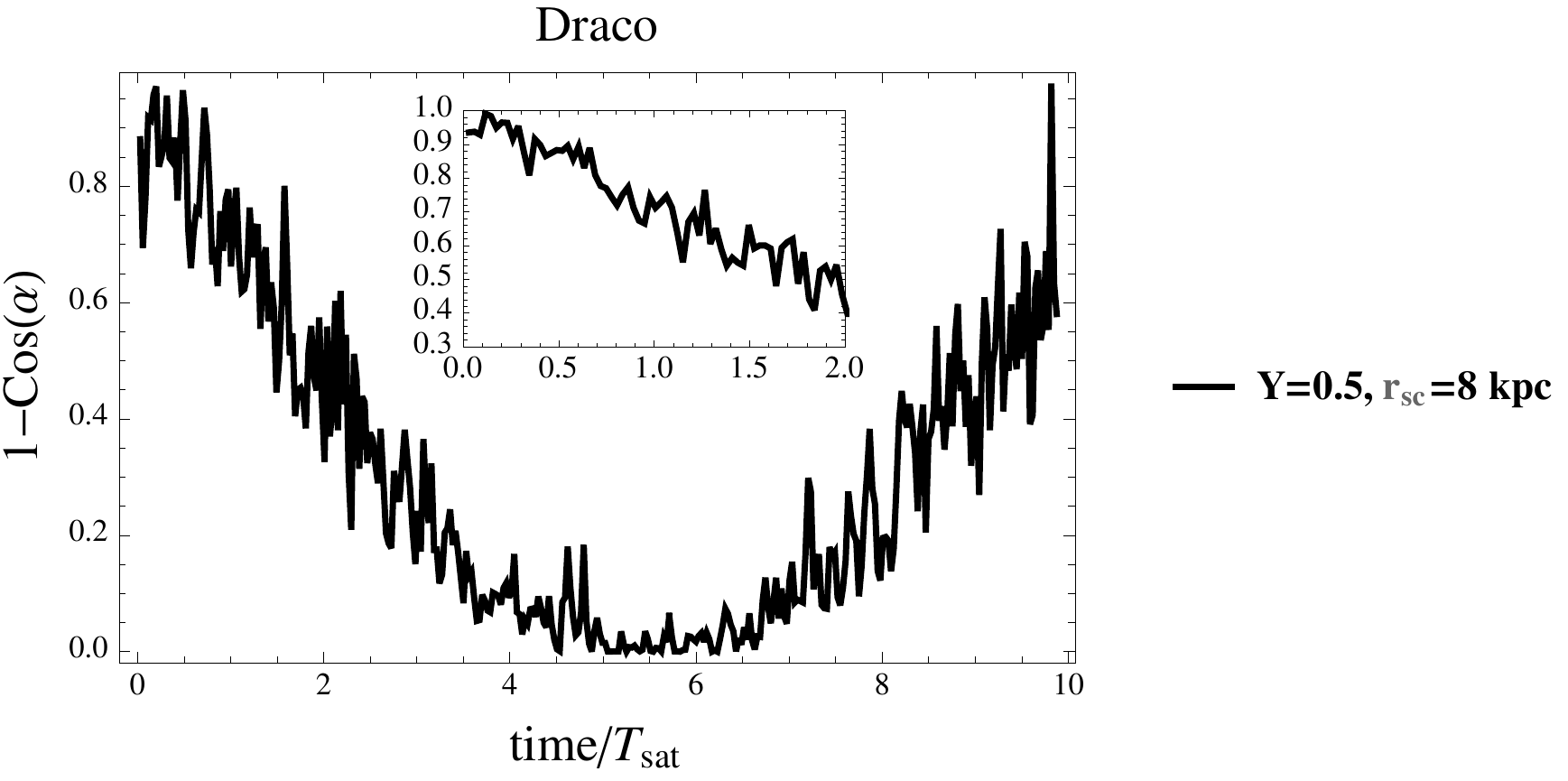}
\caption{Time evolution of the angle between the ellipticity vector
  and the one pointing towards the center of the host halo for Draco
  as a function of time measured in revolution periods 
($T_{Draco}=2.17$ Gyrs). The LV parameters are $Y=0.5$ and $r_{sc}=8$
kpc. Unity corresponds to orthogonality while zero means alignment
between the two vectors.} \label{fig:precession} 
\end{figure}

It is interesting to further examine the orientation of the major
axis with respect to the direction of motion of the satellite.  
To address this point we show in Fig.~\ref{fig:precession}
the angle between the major axis 
of Draco A and the direction to the center of the
host halo (this direction is perpendicular to the velocity of the
satellite). As advocated above, the angle changes only slightly during
the orbital period. However, it precesses significantly on longer
timescales comparable to the Hubble time.  
Using the expressions \eqref{eq:genv}, \eqref{eq:genoutv} for the LV
vector one finds that the LV force acting on a particle in the subhalo
(the term in square brackets in equation \eqref{eq:sat_p}) contains
a contribution
$$Yw'(\tilde r)\big[
\left(\tilde{\textbf v}_p\cdot \hat{\textbf r} \right)\,\tilde V_0^i
-
(\tilde{\textbf v}_p\cdot \tilde{\textbf V}_0)\,
\hat r^i\big]\;,$$
where $\hat r^i$ is the unit vector pointing from the center of the
satellite to the particle's position. By switching of and on this term
in numerical simulations we have identified it as being responsible
for the secular change in the subhalo orientation shown in
Fig.~\ref{fig:precession}.

\section{Discussion and conclusions}
\label{sec:conclusions}

We have performed  a preliminary investigation of the dynamics of
galactic satellites in the presence of a LV force in the DM
sector. This extends previous studies of the effects of LV in DM to
shorter scales. 
This is a promising arena for testing LV in the dark sector with potential observational signatures. The relevant range of parameters enters through 
the screening scale $r_{sc}$ \eqref{eq:def_rsc} and direct effects
related to the DM coupling to LV, $Y$. The typical probed range
$r_{sc}\sim $kpcs is remarkably different from that constrained from
linear dynamics at large scales, $r_{sc }\sim $ Mpcs. 
The price to pay is the complexity of modeling the highly non linear
structures whose detailed  dynamical evolution must ultimately be
followed with full numerical simulations.  
The goal of the present paper has been to identify generic signatures
of LV forces. Hence, we have opted to follow  a simplified scheme
designed to  capture the main consequences of the model. 
 Our focus have been  the dynamics of a satellite orbiting
 around its host halo. For this reason we have been concerned only
 with the effects on the satellite alone. Therefore, we have
 restricted the analysis to  the parameter regime for which the LV
 effects of the host halo are screened. Besides, we neglected tidal
 gravitational forces and left out the formation process of the
 satellite -- host halo system.

We have identified  several interesting effects which could have
substantial implications on the  
evolution of  satellite galaxies: An enhancement of the inner halo
density accompanied  with  mass extraction from the outer regions and
a distortion of the halo shape. The main features of these effects
are: 

\paragraph*{Mass profile:}

The LV force produces  a significant  compression of  the  satellite
halo. This translates into a different gravitational field with
respect to the $\Lambda$CDM model acting on the stellar component. In
this respect, a  potential  way to seek signatures of (or
constraints on) LV is  by measurements of the
velocity dispersion profile of the stars in the satellite as we have
discussed in section 
\ref{sec:constr_mass_distrib}.
We have found
that the curves obtained
within LV models differ significantly from those where no LV force is
acting. 
However, both are still compatible with the observations
cf. Fig.~\ref{fig:sigmalos_Fornax}. 
It would be interesting to explore if more precise data
\cite{Walker:2009zp} are capable to discriminate between Lorentz
invariant and LV models, which, however, requires more detailed and
realistic numerical simulations.  
 
\paragraph*{Total mass of subhalos:} A feature of the LV force is its
ability to efficiently extract matter from the outskirts of
subhalos. This  mechanism for particle extraction works in addition to
tidal stripping  by  the  gravitational force field of the host
halo. Depending on the relative importance of the two effects, various
scenarios can be envisaged. If the gravitational tidal stripping is
very efficient, we should not expect much particle extraction via LV,
as we have seen in the case of Draco A and Fornax. However, if the
opposite situation is realized, as is the case for Draco B,  the LV
yanks matter at the outer halo regions and acts on shorter timescales
than the
gravitational tidal stripping, leaving the latter largely irrelevant.  
This could result in
a different equilibrium configuration.   
Although the residual  mass from the two processes is similar, the
final halo will show very different characteristics that will allow
for a distinction between the two mechanisms.  Tidal stripping does
not greatly alter the density profile, in contrast  to the LV
force. Moreover, gravitational stripping produces symmetric streams
(leading and trailing); in the case of LV force, streams of extracted
stars would be asymmetric, reflecting the preferred direction defined
by the vector. This latter feature can be noticed in the right panels
of figure~\ref{fig:nocore_xy} where a characteristic tail is formed
after a significant amount of mass has been extracted from the
satellite and more dramatically in the second figure of the Draco B
case. 

\paragraph*{Ellipticity of the satellite halo:}

The LV force produces a distortion in the shape of  subhalos. 
We found that the ellipticity of the altered shapes is in the ballpark
of the values observed for dwarf satellites in the Local Group
\cite{2008ApJ...684.1075M,ellipticity}, but the latter are also
compatible with pure  
$\Lambda$CDM simulations
\cite{2014MNRAS.439.2863V,2015MNRAS.447.1112B}. 
More detailed numerical studies are required to tell apart the LV effect from
the purely gravitational one.
The distortion alters the gravitational potential felt by stars,
which can be directly probed  
by measurements of the stellar velocity dispersion.
Unfortunately, inferring the shape-related LV signatures is hampered  by the
precession of the major axis which smears  
the correlation of LV generated ellipticity with the direction of
motion of the satellite. 

The challenge in constraining LV effects is that standard gravity
could also be associated with similar phenomenology. Our analysis
points out several interesting features and motivates a more
sophisticated numerical and observational modeling for disentangling
LV from purely gravitational signatures. \\

As we discussed in the introduction, there is a strong motivation 
to  test Lorentz invariance. 
For low energies and late time Universe this requires the study of models with Lorentz violating fields. 
These fields are generically coupled to the other `standard'
components of the Universe, and these interactions control the effects
of LV 
in different observations. Although the couplings to the standard model particles are highly
constrained, the same is not true for DM.  Given the current and forthcoming
observational data, the investigation of the effects of LV in  DM models using small scales structures may not only reveal
interesting aspects of how such structures have formed but could also
shed light on this fundamental aspects of Nature.

\section*{Acknowledgments}
This research was supported by the I-CORE Program of the Planning and
Budgeting Committee, THE ISRAEL SCIENCE FOUNDATION (grants No. 1829/12
and No. 203/09), the Asher Space Research Institute. 
D. Bettoni acknowledges financial support form ``Fondazione Angelo
Della Riccia'' and from the SFB-Transregio TRR33 ``The Dark Universe''. D. Bettoni wish to thank Nordita where a large part of this work has been carried on.  D. Blas would like to thank the Physics Department of
the Universidad de Chile for its warm hospitality during the
completion of this work. The work of S.S. is
supported by the Swiss National Science Foundation.  

\appendix

\section{analytic solution for the LV vector}
\label{app:analytic_LV}

We give now the detailed derivation of the solutions of the equation
for the LV vector $\textbf{u}$ discussed in section
\ref{sec:LVDM_AR}. Here, we consider the general case of a spherical object (the
satellite) which has a core of constant density surrounded by a shell
where the density is dropping as $r^{-2}$ and whose particles are
collectively moving with the speed $\vV_s$. The satellite moves  inside a
bigger object (the host halo) with constant density and whose
particles are moving with constant bulk velocity $\vV_h$.  
We have to solve the following three equations
\begin{eqnarray}
\tilde{\Delta} \tilde u^i & = & \alpha^2_c \left(\tilde u^i-\tilde V^i_s\right) \,,\quad \text{if} \quad r\le r_c\,,\\
\tilde{\Delta}\tilde  u^i & = & \left(\frac{\tilde r_v}{\tilde r}\right)^2 \left(\tilde u^i-\tilde V^i_s\right) \,,\quad \text{if} \quad r_c\le r\le r_s\,,\label{eq:midh}\\
\tilde{\Delta} \tilde u^i & = & \alpha_h^2 \left(\tilde u^i-\tilde V^i_h\right) \,,\quad \text{if} \quad r \ge r_s\,,
\end{eqnarray}
where $r_v$ and $r_s$ are the virial and the actual radius of the
satellite respectively and $r_c$ is the radius of the core. The scale
$r_{sc}$ is defined in \eqref{eq:def_rsc}.
Requiring that the transition between the three regions is smooth,
i.e. $\rho_a(\tilde r_a)=\rho_s(\tilde r_a)$ for
$a=h,c$, yields  
$\alpha_a=\tilde r_v/\tilde r_a$. The solution to these
equations reads, 
\begin{eqnarray}
\tilde u_c^i(\tilde r) &=&\tilde  V^i_s+C^i_1 \frac{e^{-\alpha_c \tilde r}}{\tilde r}+ C^i_2 \frac{e^{\alpha_c \tilde r}}{\tilde r} \,,\quad \text{if} \quad r\le r_c\,,\\
\tilde u_s^i(\tilde r) &=& \tilde V^i_s+C^i_3 \,\tilde r^{n_-}+C^i_4 \,\tilde r^{n_+}\,,\quad \text{if} \quad r_c\le r\le r_s\,,\\
\tilde u_h^i(\tilde r) &=&\tilde  V^i_h + C^i_5 \frac{e^{-\alpha_h \tilde r}}{\tilde r}+C^i_6 \frac{e^{\alpha_h\tilde  r}}{\tilde  r}\,,\quad \text{if} \quad r \ge r_s\,,
\end{eqnarray}
where $n_+= (-1+\sqrt{1+4\tilde r_v^2})/2$ and $n_-=
(-1-\sqrt{1+4\tilde r_v^2})/2$.  

The solutions are required to be finite at $r=0$ and vanishing at
infinite radius. This fixes two of the six integration constants. In
particular, we get  $C_1=-C_2$ and
$C_6=0$. By matching the solutions at $r=r_c$
and at $r=r_s$ we obtain 
\begin{eqnarray}
\label{eq:coresat}
\tilde u^i_c(\tilde r)&=&\tilde V^i_s-\left(\tilde V^i_s-\tilde V^i_h\right)\frac{\tilde r_c^{n_+}-\Gamma_c \tilde r_c^{n_-}}{\tilde r_s^{n_+}\kappa_+-\tilde r_s^{n_-}\kappa_-\Gamma_c}\frac{\sinh(\alpha_c \tilde r)}{\sinh(\alpha_c \tilde r_c)}\frac{\tilde r_c}{\tilde r}\,, \quad \text{if} \quad r\le r_c\,,\\
\tilde u^i_s(r)&=& \tilde V^i_s -\left(\tilde V^i_s-\tilde V^i_h\right)\frac{\tilde r^{n_+}-\Gamma_c \tilde r^{n_-}}{\tilde r_s^{n_+}\kappa_+-\tilde r_s^{n_-}\kappa_-\Gamma_c}\,, \quad \text{if} \quad r_c\le r\le r_s\,,\\
\tilde u^i_h(r)&=&\tilde V^i_h+\left(\tilde V^i_s-\tilde V^i_h\right)\left[1-\frac{\tilde r_s^{n_+}-\Gamma_c\tilde r_s^{n_-}}{\tilde r_s^{n_+}\kappa_+-\tilde r_s^{n_-}\kappa_-\Gamma_c}\right]\frac{e^{-\alpha_h(\tilde r-\tilde r_s)}}{\tilde r/\tilde r_s}\,, \quad \text{if} \quad  r \ge r_s\,,
\end{eqnarray}
where
\begin{equation}
\kappa_\pm = 1+\frac{n_\pm}{1+\alpha_h\tilde  r_s}\,, \quad \Gamma_c=\frac{\gamma_c-n_+}{\gamma_c-n_-}\cdot\frac{\tilde r_c^{n_+}}{\tilde r_c^{n_-}}\,, \quad \gamma_c=\alpha_c \tilde r_c\coth(\alpha_c \tilde r_c)-1\,.
\end{equation}

\begin{figure}[ht!]
\centering
\includegraphics[width=0.45\columnwidth]{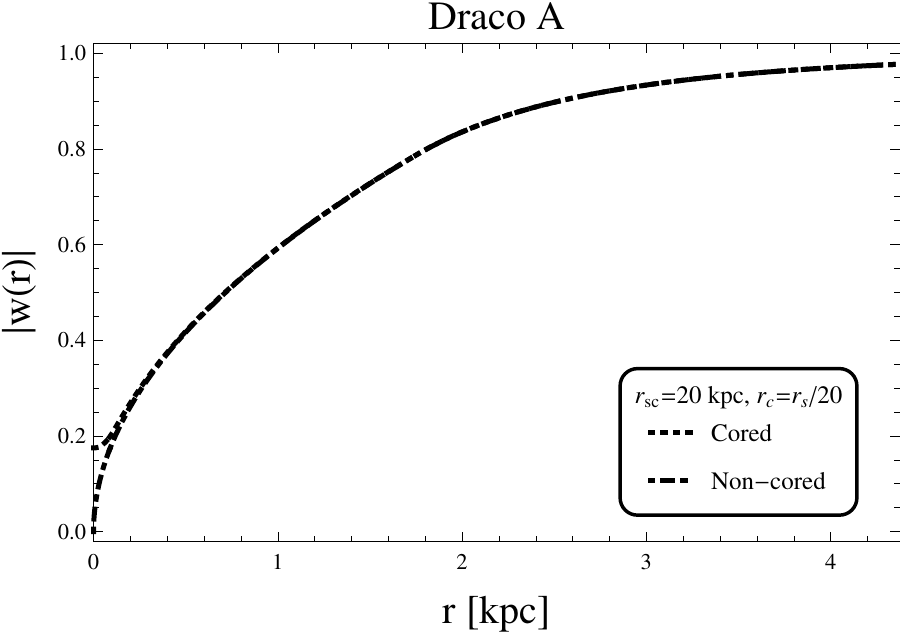}
\caption{Comparison between the analytic solutions for the LV vector
  $u^i$ obtained with a pure $\rho \sim r^{-2}$ profile and with a
  $\rho \sim r^{-2}$ profile that has a core of constant density, in
  the case of Draco A. Plotted is the radial time independent part of the LV solution, $|w|$, defined in \eqref{eq:genv} and \eqref{eq:genoutv}. The screening scale is
  $r_{sc}=20$~kpc and the virial radius is $r_v=20$~kpc.}\label{fig:cored_vs_rho2}  
\end{figure}
From this general solution we can recover the two limits that we have
used in this work, namely the inverse square law and the purely
constant profiles.  
To get the first one we  take $r_c=0$. In this case the solution is 
\begin{eqnarray}\label{app:eq:ur2}
\tilde u^i_{s}(\tilde r)&=&\tilde  V^i_s + \left(\tilde V^i_h-\tilde V^i_s\right) \frac{1}{1+\frac{n_+}{\alpha_h \tilde r_s+1}}\left(\frac{\tilde  r}{\tilde  r_s}\right)^{n_+}\,\\
\tilde u^i_{h}(\tilde r) &=&\tilde  V^i_h - (\tilde V_h^i-\tilde V_s^i)\frac{n_+}{1+n_++\alpha_h\tilde r_s}\frac{e^{-\alpha_h(\tilde r-\tilde r_s)}}{\tilde r/\tilde r_s}\,.
\end{eqnarray}
 The second limit is obtained by  $r_c\rightarrow r_s$ and defining $\alpha=\alpha_h/ \alpha_c$, yielding  the solution
\begin{eqnarray}\label{app:eq:urc}
\tilde u^i_{s}(\tilde r)&=& \tilde V^i_s + \left(\tilde V^i_h-\tilde V^i_s\right)\left[\tilde r_s+\frac{\alpha_c^{-1}\sinh(\alpha_c \tilde r_c)-\tilde r_c \cosh(\alpha_c\tilde r_c)}{\alpha \sinh(\alpha_c\tilde r_c)+\cosh(\alpha_c\tilde r_c)}\right]\frac{\sinh(\alpha_c \tilde r)}{\sinh(\alpha_c\tilde r_c)}\frac{1}{\tilde r}\,\\
\tilde u^i_{h}(\tilde r) &=& \tilde V^i_h + (\tilde V_h^i-\tilde V_s^i)\frac{\alpha_c^{-1}\sinh(\alpha_c \tilde r_c)-\tilde r_c\cosh(\alpha_c\tilde r_c)}{\alpha \sinh(\alpha_c\tilde r_c)+\cosh(\alpha_c\tilde r_c)}\frac{e^{-\alpha_h(\tilde r-\tilde r_s)}}{\tilde r}\,.
\end{eqnarray}

\begin{figure}[ht!]
\centering
\includegraphics[width=0.45\columnwidth]{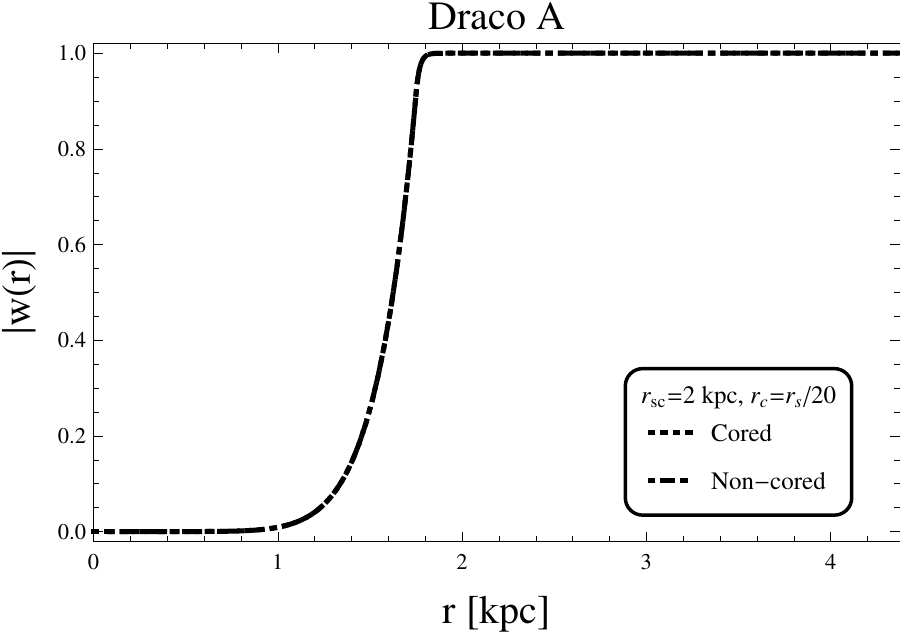}
\caption{Same as Fig.~\ref{fig:cored_vs_rho2}, but for $r_{sc}=2$~kpc and the virial radius is $r_v=20$~kpc. In this case the two curves coincide.
} \label{fig:cored_vs_rho2_screen}
\end{figure}

In figures \ref{fig:cored_vs_rho2} and \ref{fig:cored_vs_rho2_screen}
we show solutions for $\omega(r)$ (as defined in \eqref{eq:genv} and
\eqref{eq:genoutv}) obtained from an inverse square law 
matter distribution for Draco A with and without a core for
$r_{sc}=20$ and $r_{sc}=2$ respectively.  
In the first case there is a slight difference between the two curves
in the center of the satellite, whereas in the second case the curves
coincide. In both cases the LV force acting on particles inside the core is
suppressed. This last point is exploited in the definition
of the satellite reference frame used for the numerical formulation of
the problem. In fact, we can consider this inner region as
screened and consequently following the standard Newtonian motion. This
in turns defines the reference frame in which the dynamics of
particles in the subhalo is governed by the equation 
\eqref{eq:sat_p}. 
In the second case $r_{sc}\le r_s$ and the LV vector solution is
constant almost up to the edge of the satellite as can be seen
in figure \ref{fig:cored_vs_rho2_screen}. This is a consequence of the
screening mechanism characteristic of the LV model. 

\bibliographystyle{JHEP}
\bibliography{mybib.bib}
\end{document}